\documentclass[letterpaper,journal]{IEEEtran}
\usepackage{amsmath,amssymb,amsfonts}
\usepackage{array}
\usepackage[caption=false,font=normalsize,labelfont=sf,textfont=sf]{subfig}
\usepackage{textcomp}
\usepackage{stfloats}
\usepackage{url}
\usepackage{verbatim}
\usepackage{graphicx}
\usepackage{cite}
\usepackage{booktabs}
\usepackage{xcolor}
\usepackage[
  colorlinks=true,
  linkcolor=blue,
  citecolor=blue,
  urlcolor=blue
]{hyperref}
\makeatletter
\def\@cite#1#2{\textcolor{blue}{[{#1\if@tempswa , #2\fi}]}}
\makeatother
\begin{document}

\title{\fontsize{15}{19}\selectfont Embodied Semantic Communication for Collective Autonomous Agents:\\
A Tutorial on Representation, Wireless Delivery, and Closed-Loop Coordination}

\author{Yizheng~Huang,~\IEEEmembership{Student Member,~IEEE}, Wensheng~Lin,~\IEEEmembership{Member,~IEEE}, Lixin~Li,~\IEEEmembership{Member,~IEEE},\\
Qinghe~Du,~\IEEEmembership{Member,~IEEE}, Wenchi~Cheng,~\IEEEmembership{Senior Member,~IEEE}, and Zhu~Han,~\IEEEmembership{Fellow,~IEEE}%
\thanks{Yizheng Huang, Wensheng Lin, and Lixin Li are with the School of Electronics and Information, Northwestern Polytechnical University, Xi'an, Shaanxi 710129, China (e-mail: hyzzz@mail.nwpu.edu.cn; linwest@nwpu.edu.cn; lilixin@nwpu.edu.cn).
\par Qinghe Du is with the School of Information and Communications Engineering, Xi'an Jiaotong University, Xi'an 710049, China (e-mail: duqinghe@mail.xjtu.edu.cn).
\par Wenchi Cheng is with the School of Telecommunications Engineering, Xidian University, Xi'an 710071, China (e-mail: wccheng@xidian.edu.cn).
\par Zhu Han is with the Department of Electrical and Computer Engineering, University of Houston, Houston, TX 77004, USA (e-mail: hanzhu22@gmail.com).}%
}

\maketitle
\bstctlcite{IEEEtranBSTCTLNoDash}

\begin{abstract}
As autonomous systems and embodied intelligence enter the dynamic physical world, multi-agent collaboration calls for a paradigm shift in communication design. However, existing communication paradigms overlook that agents form action understanding from their own states, environmental observations, and collaboration relations through a process that evolves as a task unfolds. Consequently, reliable bit delivery, general semantic recovery, or single-task utility optimization alone cannot ensure that heterogeneous agents form coordinated actions compatible with their own conditions from shared information during task execution. To address this gap, this paper proposes embodied semantic communication (ESC) as a paradigm that transforms information transmission into action-oriented semantic interaction. Specifically, ESC characterizes how an explicit communication link can encapsulate multimodal perceptual states, intrinsic hardware capabilities, and collaborative intents into unified actionable semantic representations, thereby enabling heterogeneous receiving agents to parse, align, and ground them in local motor control. This paper clarifies the conceptual boundary, system characteristics, and environment-constrained technical pathways of ESC. It maps the underlying mathematical tools, including semantic information theory, world models, and multi-agent decision theory. Finally, this paper summarizes key open challenges, including measurable semantic reliability, ambiguity-triggered interaction under dynamic environments and tasks, and bandwidth-adaptive semantic transmission, outlining a roadmap for collective embodied networks.
\end{abstract}


\section{Introduction}

\subsection{Background and Motivation}

In recent years, embodied intelligence has undergone rapid paradigm expansion, driving autonomous agents from idealized virtual test environments into uncertain physical ecosystems to perform complex, long-horizon interactive tasks~\cite{Sun2024EmbodiedIntelligenceSurvey,MonWilliams2025EmbodiedLLMRobot}. This evolution essentially requires multimodal perception, cognitive decision-making, and physical motor execution to be tightly and continuously coupled in uncertain environments. More importantly, as task scales exceed the capability boundary of a single entity and push systems toward collective collaborative intelligence, the core objective of wireless networks also changes fundamentally. A network is no longer only a transmission conduit that connects static topologies, homogeneous data pipelines, or abstract terminal devices. Instead, it must dynamically interconnect heterogeneous embodied agents with different physical constraints, sensing capabilities, and action capabilities, so that they can achieve closed-loop collaboration through environmental interaction~\cite{Liang2026WirelessEmbodiedIntelligence}.

In multi-agent collaboration oriented embodied tasks, communication is no longer limited to bit-level information transmission. It increasingly becomes a collaborative mechanism that connects perception, decision-making, and action. In complex scenarios such as multi-agent collaborative reconnaissance and vehicle-to-everything (V2X) autonomous transportation, a single agent is always constrained by limited sensing range, its own kinematic limitations, and finite onboard resource budgets. Therefore, complex tasks require an explicit integration of local closed-loop control at the individual level and distributed macro-level coordination at the collective level~\cite{Petersen2019CollectiveRoboticConstruction}. Recent studies on constraint-driven topologies further show that distributed information-exchange architectures fundamentally determine how robustly a swarm adapts to spatiotemporal environmental fluctuations~\cite{Talamali2021WhenLess}. This observation reveals a deeper requirement: the communication demands of next-generation collective embodied networks must go beyond Shannon-style error-free propagation and integrate cooperative semantic perception, distributed decision feedback, real-time motion coordination, and resource-adaptive semantic selection as a whole.

Cooperative perception originates from the spatiotemporal locality inherent in the observation perspective of a single agent. Conventional cooperative perception schemes in autonomous transportation usually rely on raw-data or feature-level telemetry exchange to mitigate individual blind spots~\cite{R028,R029}. \emph{Embodied semantic communication (ESC)}, which communicates action-relevant embodied semantics rather than raw observations, shifts the operational focus from large-scale data replication to the extraction of environmental semantics that can be reused and adapted by multiple agents. Raw video streams, high-density point clouds, and uncompressed multimodal streams impose prohibitive wireless overhead and widespread informational redundancy~\cite{Ma2026Condensed360,Ma2026LargeAIImmersive}. Under the ESC paradigm, sharing is optimized as semantic transmission oriented toward structural abstraction~\cite{Feng2025SemanticAutonomousDriving}. For example, in a multi-robot warehouse fleet, when an agent is passing through a blocked corridor, it does not transmit its raw visual telemetry. Instead, it synthesizes and broadcasts a high-level semantic proposition, such as ``left corridor: temporarily occluded by a moving asset''. This semantic information enables the following units to replan their paths in advance or adjust their velocity profiles accordingly.

Beyond perceptual extension, decision assistance under the ESC framework also requires the transmitted semantics to be deeply embedded into closed-loop processes such as task-state estimation, risk assessment, and motion-behavior selection. This requirement can be seen more intuitively from human collaboration. During a joint operation, a simple instruction may trigger multiple rounds of asymmetric linguistic feedback to resolve spatial-reference ambiguity or to dynamically adjust the execution strategy when the operating location changes. In a warehouse picking task, for instance, worker A may ask worker B to deliver the ``blue component on the shelf'' to the assembly station. If the target type is unique, B can act directly. If multiple candidate targets or target types exist, B needs to ask for the model number. If the assembly-station location changes, subsequent communication shifts from ``which object to take'' to ``where to deliver it''. Transferring this interaction mechanism to collective embodied networks requires a fundamental departure from traditional one-shot, open-loop semantic transmission. Communication can no longer be modeled as a static and decoupled process. Instead, it must co-evolve with task progress over the entire task horizon, become a dynamic state-update protocol, and be explicitly optimized for ambiguity resolution and real-time behavior adjustment under structured task uncertainty.

\begin{figure*}[!t]
  \centering
  \includegraphics[width=0.99\textwidth]{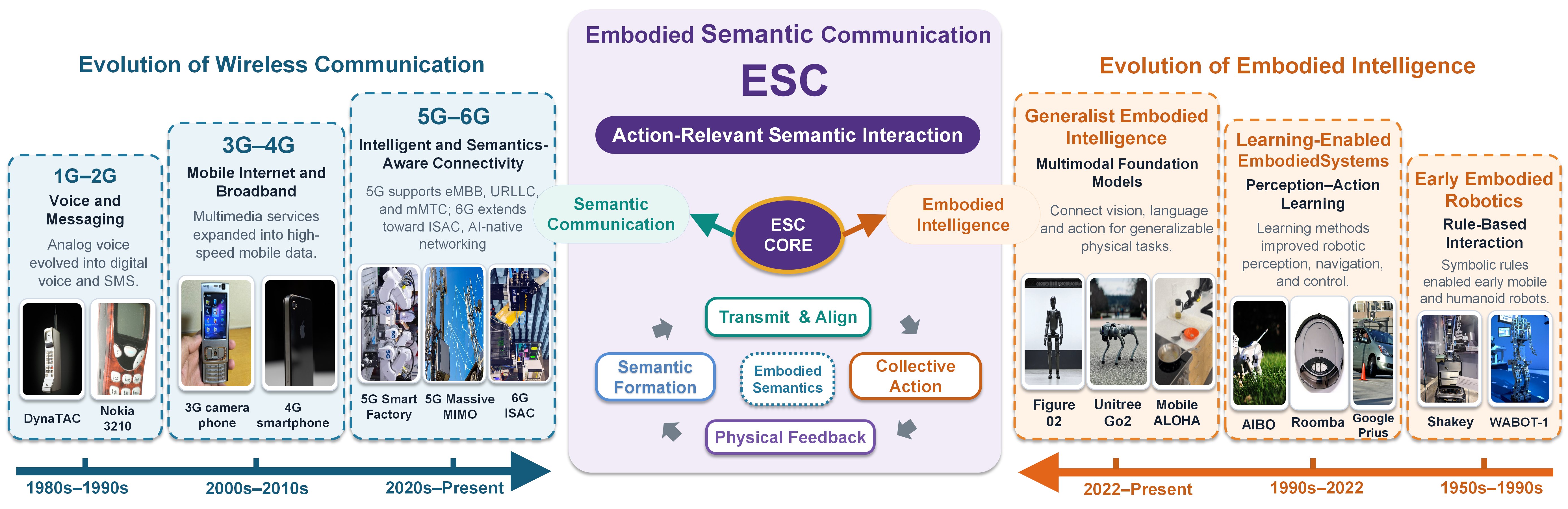}
  \caption{Convergence of wireless communication and embodied intelligence toward embodied semantic communication. Wireless communication evolves from voice and messaging toward intelligent and semantics-aware connectivity, while embodied intelligence progresses from rule-based robotics to learning-enabled and foundation-model-driven systems. At their convergence, ESC organizes semantic formation, transmission and alignment, collective action, and physical feedback around embodied semantics.}
  \label{fig:esc_convergence}
\end{figure*}
These requirements also clarify why embodied semantics cannot be reduced to the conventional ``\emph{shorter is better}'' principle of semantic compression. ESC carries two complementary forms of information: coordination-oriented control signaling and unified perceptual semantics reusable across tasks. For control signaling, effectiveness depends less on descriptive completeness than on whether the receiver can combine sparse cues with shared context and interaction history to recover operational intent. Compression therefore removes explicit content that the receiver can reliably reconstruct from its own context, rather than indiscriminately discarding task-relevant meaning.

For unified perceptual semantics, ESC also differs from semantic communication designed for deterministic tasks. Many current studies on task-oriented semantic communication usually predefine the downstream task, such as classification, detection, or segmentation. In collective embodied tasks, however, different receivers often do not have a single task. The same perceptual information may serve different decision processes, such as local motion adjustment and task-role allocation. Therefore, the sender cannot provide only the result of a single task. It needs to transform multimodal observations into unified semantics reusable by different receivers. Such semantics should remove ineffective redundancy across modalities while retaining necessary task redundancy according to communication resources. Jia \emph{et al.}~\cite{R030} show, in raw-level cooperative perception scheduling, that limited wireless bandwidth requires sensing sharing to be organized around task and receiver needs. Studies by Lin \emph{et al.}~\cite{R031} and Liu \emph{et al.}~\cite{R064} on vehicle-to-vehicle cooperative perception also reflect the trade-offs among raw-data sharing, feature sharing, and result sharing in terms of bandwidth, computation, and perception performance. At the same time, the semantic relaying framework proposed in~\cite{Lin2024SemanticForwardRelaying}, in which a relay node extracts semantic features with side information and forwards them as a more compact communication object, further illustrates the shift from single-task results or raw packets to reusable semantic representations. Therefore, compression in ESC is not simply reducing data volume. It dynamically selects semantic granularity by balancing receiver task uncertainty, heterogeneous needs, and communication resources, deciding which semantics must be transmitted, which information can be compressed, and which content can be completed by the receiver using context.

Although frontier studies in collective intelligence provide valuable foundational modules for related problems, they cannot fundamentally replace the ESC paradigm. For example, communication protocols in conventional multi-agent deep reinforcement learning (MADRL) usually maximize accumulated team rewards or accelerate joint-policy convergence by optimizing continuous message interaction~\cite{Zhu2024CommMADRL}. However, such methods essentially rely on uninterpretable black-box latent-vector exchange and remain fragile under practical non-ideal wireless physical-layer conditions. Some recent pioneering works have begun to discuss semantic-distortion mitigation in multi-robot systems, but they often treat semantic components as local modules appended to existing communication topologies~\cite{Chen2026TaskDrivenSemanticCollaborativeCommunication}. In contrast, the key issue in ESC is \emph{not to add a semantic module to an existing multi-agent communication framework}. It is to reorganize communication content under the needs of collective embodied intelligence, so that communication effectiveness is ultimately grounded in collective embodied closed-loop action.

\subsection{Definition and Scope}

ESC lies at the convergence of the two technological trajectories of wireless communication and embodied intelligence, as shown in Fig.~\ref{fig:esc_convergence}. As represented by successive cellular generations, wireless communication has evolved from voice and messaging in 1G--2G, through mobile Internet and broadband in 3G--4G, toward intelligent and semantics-aware connectivity in 5G and prospective 6G systems. Semantic communication has emerged alongside AI-native networking and integrated sensing and communication as an important direction for future wireless systems. At the conceptual level, Shannon's mathematical theory established rigorous limits for reliable symbol transmission over noisy channels~\cite{R001}, while Weaver's classical three-level distinction anticipated the need to address meaning and behavioral influence~\cite{R002}. Subsequent goal- and task-oriented communication frameworks introduced operational utility~\cite{R003,R006}, whereas semantic communication shifted the focus from bit-level replication to the expression and reconstruction of meaning~\cite{R005}. In parallel, embodied intelligence has progressed from rule-based interaction to learning-enabled perception--action systems and, more recently, multimodal foundation-model-driven agents.

ESC emerges where these two trajectories converge. This paper formalizes embodied semantics as action-related informational meaning jointly constrained by an agent's physical state, environmental observations, task objective, local risks, and collaboration relations. When such meaning is formed, transmitted, parsed, and aligned through explicit wireless links to support collaborative behavior, it constitutes ESC. As illustrated by the central loop in Fig.~\ref{fig:esc_convergence}, ESC extends beyond semantic recovery by integrating semantic formation, transmission and alignment, collective action, and physical feedback throughout a task cycle, thereby enabling heterogeneous receiving agents to convert shared semantics into actions according to their own embodiment.

Specifically, this paper gives the following formal system-level definition of ESC: in collective embodied-intelligence scenarios, ESC is a specialized communication paradigm in which a sending agent encapsulates its high-dimensional embodied attributes, including multimodal perceptual states, intrinsic kinematic capabilities, operational intents, and structured collaborative relations, into compact and transmissible semantic representations, and adapts them to conventional or next-generation physical-layer channels~\cite{Fu2024DigitalAnalog}. After reception, heterogeneous receiving agents dynamically parse and align these semantic representations with their own local exteroceptive constraints and proprioceptive states, so as to execute collaborative decision-making while minimizing latency and safety risk and maximizing collective task efficiency. Crucially, ESC is not an isolated information exchange. Instead, it is directly embedded in the \emph{closed-loop physical action-feedback cycle} of collective intelligence.

ESC transmits a unified semantic representation constrained by action conditions. It is mainly synthesized from two informational dimensions: \emph{multimodal perceptual semantics} and \emph{distributed task intent}. For perceptual semantics, ESC focuses on mapping heterogeneous source observations, such as images, speech, and tactile signals, into a semantic space reusable by multiple agents, so that heterogeneous receivers can align the received semantics with their own environmental understanding and task judgment. Compared with conventional distributed sensing systems that directly extend sensing boundaries, ESC emphasizes reducing ineffective redundancy across modalities and nodes while extending the perceptual range. For task intent, ESC emphasizes using shared semantics to support distributed collaboration rather than direct, fine-grained command decomposition. For example, in a cooperative object-transport task, the initiating agent only needs to transmit a single abstract task proposition, namely ``execute cooperative lifting''. After receiving this high-level intent, nearby heterogeneous units combine their own positions, load capacities, and observations of peers to locally form strategies for posture adjustment, torque balancing, and synchronized force application, thereby reducing the overhead of explicit motion signaling.

At the same time, the communication process of ESC can be characterized as \emph{embodied-semantic updating over a task cycle}. Different communication rounds may serve perception completion, ambiguity resolution, intent synchronization, or risk correction. Its effectiveness depends not only on link reliability, but also on whether the receiver correctly forms action understanding that matches its own state. Therefore, the evaluation of ESC should jointly consider semantic parsing and alignment, collective task-completion quality, collaboration efficiency, resource and latency costs, and safety-risk control. Table~\ref{tab:comm_paradigm_comparison} contrasts the three paradigms in terms of communication objective, interpretation basis, feedback role, and evaluation criterion.

\subsection{Contributions and Organization}

Table~\ref{tab:related_surveys} summarizes the differences between this survey and existing related reviews. Most existing reviews focus on the representation and recovery of semantic information, or on how AI enhances specific functions and applications within communication systems. The fundamental distinction of this survey is that it treats action-relevant embodied semantics as the object of communication and develops ESC around the transmission, parsing, and alignment of such semantics among heterogeneous autonomous agents, so that communication outcomes directly support collective embodied action loops.

\begin{table*}[!t]
\caption{Comparison of three communication paradigms.}
\label{tab:comm_paradigm_comparison}
\centering
\includegraphics[width=0.7\textwidth]{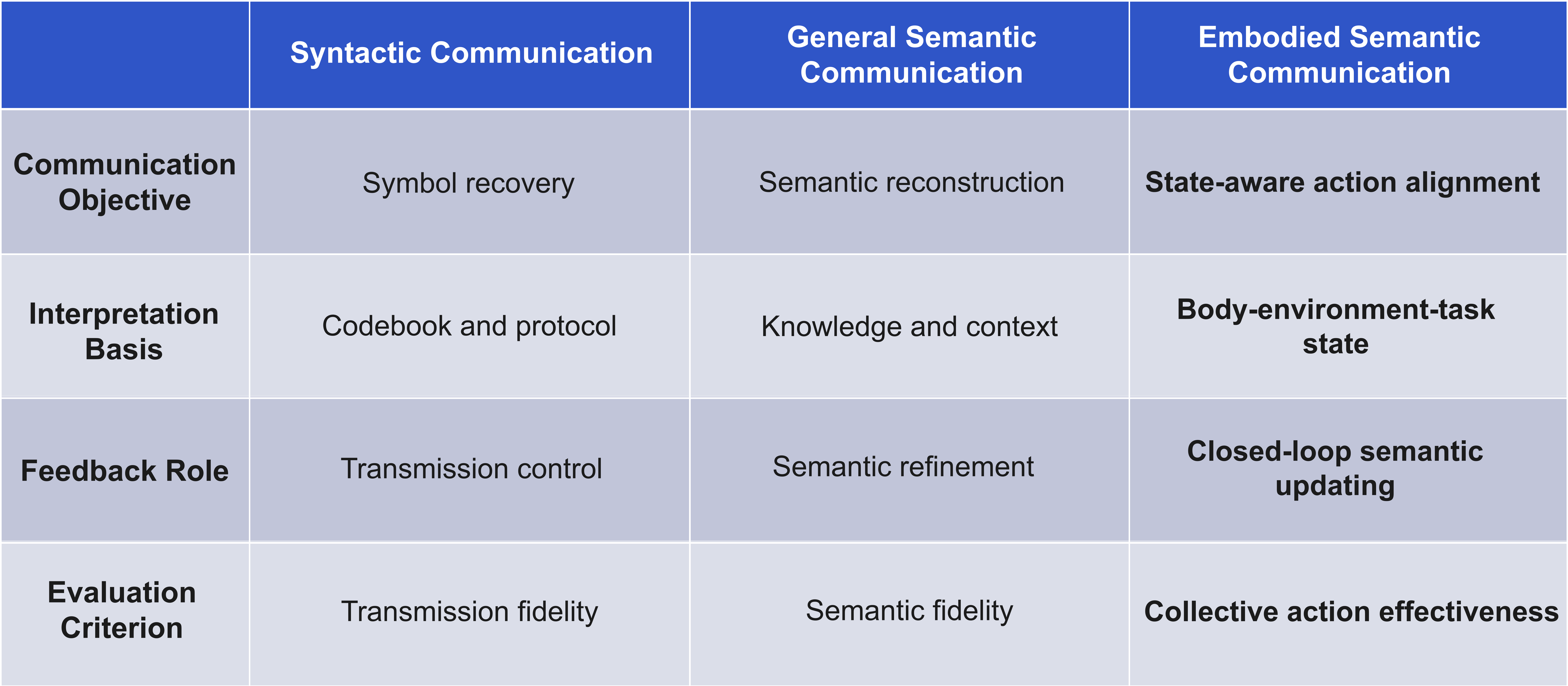}
\end{table*}

\begin{table*}[!t]
\caption{Comparison of representative surveys.}
\label{tab:related_surveys}
\centering
\scriptsize
\setlength{\tabcolsep}{4pt}
\renewcommand{\arraystretch}{1.28}
\begin{tabular}{|>{\raggedright\arraybackslash}m{0.12\textwidth}|>{\centering\arraybackslash}m{0.045\textwidth}|>{\raggedright\arraybackslash}m{0.20\textwidth}|>{\raggedright\arraybackslash}m{0.535\textwidth}|}
\hline
\textbf{Reference} & \textbf{Year} & \textbf{Primary Focus} & \textbf{Main Contributions} \\
\hline
Lu \emph{et al.}~\cite{Lu2024SemanticsEmpoweredSurvey}
& 2024
& General semantic communication
& Reviews architectures, enabling technologies, evaluation metrics, applications, and open issues in semantic communication. \\
\hline

Qin \emph{et al.}~\cite{R004}
& 2024
& AI-empowered wireless communications
& Presents an overview of AI/ML-empowered wireless communications at the physical and lower MAC layers and AI/ML-enabled semantic communication systems. \\
\hline

Wu \emph{et al.}~\cite{Wu2025AIEnabledISCCSurvey}
& 2025
& AI-enabled integrated sensing, communication, and computation
& Reviews key technologies, system architectures, evaluation metrics, and AI integration for ISCC systems. \\
\hline

Chaccour \emph{et al.}~\cite{Chaccour2025LessDataMoreKnowledge}
& 2025
& Knowledge- and reasoning-driven semantic networks
& Proposes an end-to-end framework for semantic representations, semantic languages, causal reasoning, knowledge accumulation, and reasoning-oriented performance measures. \\
\hline

Zhang \emph{et al.}~\cite{Zhang2025IntelliciseWirelessSurvey}
& 2025
& Intellicise wireless networks
& Presents a framework for intellicise wireless networks based on semantic communication. \\
\hline

Liang \emph{et al.}~\cite{Liang2025GenerativeAISemanticSurvey}
& 2025
& Generative-AI-driven semantic communication
& Reviews transceiver design, semantic-effectiveness evaluation, knowledge management, network management, and applications of generative-AI-driven semantic communication. \\
\hline

Khoramnejad and Hossain~\cite{Khoramnejad2025GenerativeAIOptimization}
& 2025
& Generative AI for wireless-network optimization
& Reviews generative models, resource allocation, network optimization, networking requirements, and open challenges for next-generation wireless networks. \\
\hline

Zhang \emph{et al.}~\cite{Zhang2026ComAI}
& 2026
& Convergence of communication and AI
& Discusses the convergence of communication and AI in terms of foundational theories, technical frameworks, experimental validation, applications, and scalability. \\
\hline

Zhu \emph{et al.}~\cite{Zhu2026CognitiveDigitalTwinSurvey}
& 2026
& Semantic communication for cognitive digital twins
& Reviews semantic codecs, network management, cognitive interaction, applications, and future directions for cognitive digital twins. \\
\hline

\textbf{Ours}
& \textbf{2026}
& \textbf{Embodied Semantic Communication}
& \textbf{Proposes ESC as a communication paradigm for collective autonomous agents and establishes a unified framework for embodied-semantic transmission, parsing, and alignment, including system characteristics, technical pathways, modeling methods, and open challenges.} \\
\hline
\end{tabular}
\end{table*}

Against the backdrop of the above definition and scope, the core contributions of this paper can be summarized in three aspects:

\begin{enumerate}
  \item Formalized concept and boundary: This paper establishes a clear conceptual ontology of ESC at the still underexplored intersection of embodied intelligence and semantic communication. By shifting the essence of communication from passive symbol recovery toward action-oriented semantic understanding, it explicitly identifies semantic transmission, semantic parsing, and multi-agent alignment as the indispensable conceptual core of ESC.
  \item System architecture construction and technical-pathway review: This paper shifts the research thread from general semantic reconstruction toward closed-loop collaboration for collective physical tasks. It systematically distills five cornerstone characteristics of ESC, including body-state semantics and environment-mediated semantics, and uses them as structural pivots to organize subsequent technical pathways, such as cross-modal compensation and cooperative perception, together with related mathematical tools.
  \item Open challenges and future directions: This paper further discusses key research challenges, including semantic reliability measurement, ambiguity-triggered interaction, and bandwidth-adaptive unified semantics. By relating these challenges to the system characteristics and technical pathways of ESC, this paper further identifies future research directions toward verifiable and deployable collective embodied networks.
\end{enumerate}

\begin{figure*}[!t]
  \centering
  \includegraphics[width=0.7\textwidth]{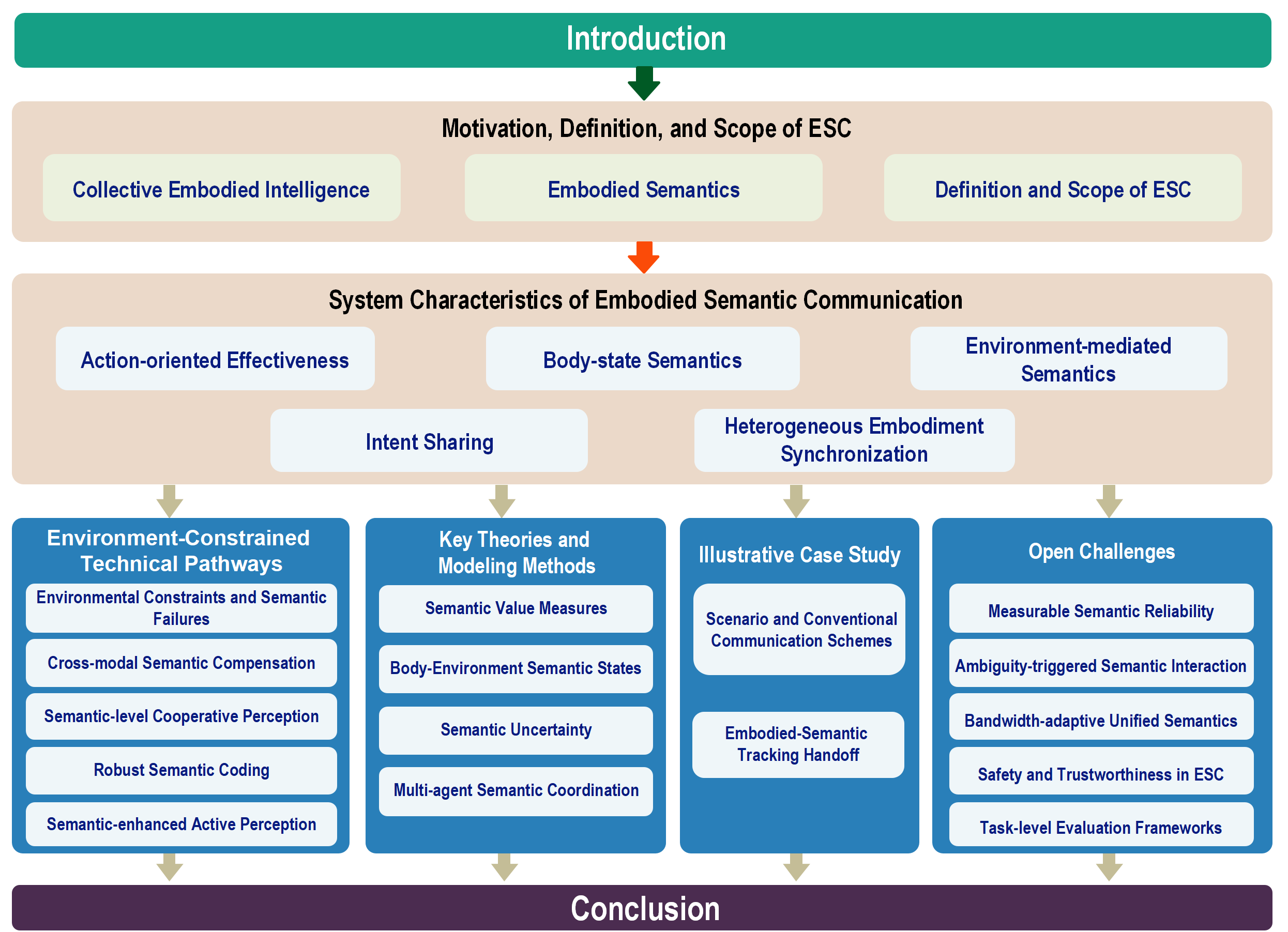}
  \caption{The structure of this paper.}
  \label{fig:article_structure}
\end{figure*}
The structure of this paper is shown in Fig.~\ref{fig:article_structure}. The remainder of this paper is organized as follows. Section~\ref{sec:system_characteristics} analyzes the intrinsic system characteristics of ESC. Section~\ref{sec:technical_pathways} reviews environment-constrained technical pathways. Section~\ref{sec:modeling_methods} develops foundational theoretical modeling methods. Section~\ref{sec:case_study} presents an illustrative case study. Section~\ref{sec:open_challenges} summarizes important open challenges and key future directions.

\section{System Characteristics of ESC}
\label{sec:system_characteristics}

For collective embodied-intelligence systems, communication effectiveness has moved beyond the evaluation frame of whether the receiver can completely reconstruct the transmitted semantic object. Although bit-level transmission reliability and semantic fidelity remain indispensable prerequisites, \emph{the optimization direction of ESC must be fundamentally reshaped around the perception--execution coupled loop}, and must further consider how communication outcomes affect task execution, multi-agent collaboration, and safety risks.

\begin{figure*}[!t]
  \centering
  \includegraphics[width=0.9\textwidth]{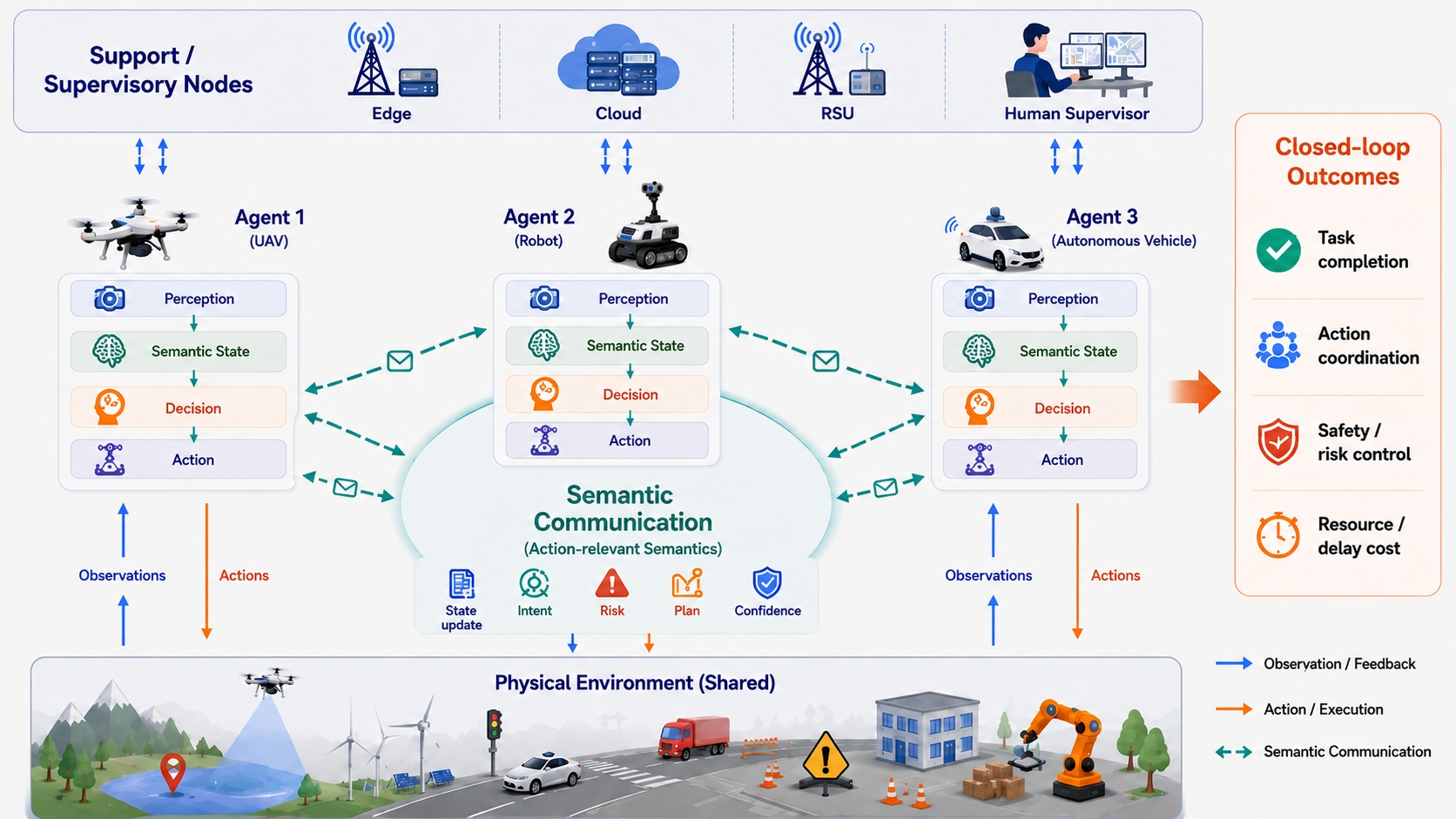}
  \caption{Closed-loop framework of embodied semantic communication. Heterogeneous agents encode local observations, body states, task intents, and collaboration needs into action-relevant embodied semantics and exchange them through explicit communication links, with optional assistance from network and human-supervisory nodes. Each receiver aligns the received semantics with its own state, task role, and local environment to update decisions and actions, while physical execution and environmental feedback continuously reshape subsequent semantic communication throughout the task cycle.}
  \label{fig:esc_closed_loop}
\end{figure*}

This tightly coupled loop architecture is shown in Fig.~\ref{fig:esc_closed_loop}. Multiple embodied agents first transform their own observations, task intents, and collaboration needs into communicable embodied semantics and transmit them among heterogeneous agents. The receiver parses and aligns the semantic information according to its own state, local environment, and task role, and then uses physical action together with environmental feedback to further influence subsequent communication needs. Around this continuous loop, this section analyzes the system characteristics of ESC from five structural pillars.

\subsection{Action-Oriented Effectiveness}
\label{subsec:action_effectiveness}

Classical communication networks mainly use bit error rate (BER), signal-to-noise ratio (SNR), and channel capacity to characterize link reliability. Semantic communication further introduces metrics such as word error rate (WER), bilingual evaluation understudy (BLEU), and semantic similarity, shifting evaluation toward semantic expression and recovery~\cite{R039,R040}. Task-oriented communication emphasizes that communication serves goal achievement, moving evaluation from semantic consistency to task benefit~\cite{R066,Fu2026GenerativeAITasCom}. Although some studies increasingly advocate connecting communication effectiveness with mathematical characterization of action outcomes~\cite{R007}, collective embodied collaboration under the ESC framework requires a more fundamental paradigm shift: communication effectiveness should no longer converge only to static task metrics, but should further examine whether communication improves group-level action outcomes, collaboration quality, and risk control.

Reference~\cite{Li2025EmbodiedMultiAgentSystems} shows that the overall performance of a collaborative system depends not only on individual capabilities, but also on coordination and control quality. Reference~\cite{KA2024MRTAReview} further includes action quality, resource consumption, and collaboration constraints in tasks such as coverage, path planning, task allocation, and object transport. For heterogeneous multi-agent systems, differences in body capability, state space, and action space also require communication to support role assignment, action connection, and collaboration rhythm~\cite{R054}. Thus, action-oriented effectiveness extends communication evaluation from single task outputs to action consequences and collaboration quality in multi-agent settings.

From the perspective of embodied-loop evaluation, semantic distortion must be understood as action-related distortion. Some studies on semantic communication have modeled semantic errors by weighting information importance according to task objectives~\cite{R041}, and state-update frameworks no longer measure value only by temporal freshness, but evaluate information value from the perspective of application-layer objective optimization~\cite{R068,Xiao2026AoGI}. ESC further introduces strict spatiotemporal coupling, meaning that semantic deviations with the same structural magnitude must be evaluated according to the physical consequences that they may induce along subsequent action chains. For example, a bounded error in target-position coordinates may be only a minor semantic difference in an environmental description task. Once this information deviation propagates into closed-loop robotic grasping, dynamic obstacle avoidance, or cooperative transport, however, it may be nonlinearly amplified and directly converted into severe grasp failure, collision, or collaboration instability. As emphasized in~\cite{R042}, distortion criteria must be dynamically adjusted according to application scenarios and task constraints. This also shifts ESC evaluation from passive semantic similarity toward active physical action consequences.

\subsection{Body-State Semantics}
\label{subsec:body_state_semantics}

Although action-oriented effectiveness requires communication outcomes to be tightly integrated into physical execution loops, this transition also requires the receiver to evaluate the execution compatibility of incoming semantics. For example, the same semantic payload, such as ``go to the second-level platform'', may have completely different action meanings for heterogeneous entities: for a multi-legged robot, it may indicate a feasible navigation path; for a wheeled platform, it may indicate an impassable constraint that requires collaborative intervention or task reallocation. Therefore, communication content cannot describe only the external environment and task goals. It must also include body states, resource conditions, and execution constraints, so that the receiver can parse embodied semantics according to its own conditions. As pointed out in~\cite{Jamone2018AffordancesSurvey}, an agent's cognition of the environment is fundamentally modulated by its action space. Robotic systems therefore need to map environmental features into local action affordances. This theoretical requirement indicates that body-state semantics must rise from local variables inside motion controllers to information content integrated into ESC encoding protocols.

This paper structurally divides body-state semantics into three levels: \emph{exteroception}, \emph{proprioception}, and \emph{interoception}. Exteroceptive environmental semantics formalize external task conditions extracted by the agent's sensors, describing information such as spatial topology, environmental features, and target features. Proprioceptive semantics describe the agent's own physical morphology and spatial dynamic state, including motion posture and body configuration. Interoceptive semantics focus on underlying system resource states, such as remaining computational resources and energy.

Exteroceptive environmental semantics can become an action basis only after being mapped to the morphological feasibility of the executing agent. External sensing information from cameras, radar, and related sensors can be extracted as task-critical semantics, such as targets, environmental features, and spatial structures, helping collective embodied systems form shared environmental understanding~\cite{R071}. However, the same environmental semantics does not necessarily correspond to the same action feasibility. A narrow passage may be a directly traversable path for a small wheeled robot, but may constitute a collision-risk constrained space for a wide load-carrying platform. Proprioception provides body-level correction in such judgment. As proposed by the authors of~\cite{R072}, collaborative fusion of exteroception and proprioception is a key condition for transforming environmental descriptions into executable action judgments. Therefore, in the ESC framework, environmental semantics must be jointly transmitted, parsed, and aligned with the agent's body state to form actionable feasibility judgment.

Meanwhile, interoceptive resource-state semantics determine the long-term sustainability of task-allocation schemes. In decentralized cooperative payload-transport tasks, a single robot node may have the dynamic capability required to fix a load, but may still be unable to support long-distance movement or sustained collective execution because of insufficient remaining onboard energy. Conventional consensus-based task-allocation methods explicitly consider energy constraints and optimize task allocation by balancing information-exchange overhead and local physical energy consumption~\cite{R073}. ESC further elevates the importance of this operating variable: as a core part of body-state semantics, interoceptive semantics can actively regulate semantic-communication selection and runtime task feasibility. Moreover, related studies on collective-intelligence task allocation jointly show that member capability differences, environmental disturbances, and capability degradation all affect role selection and task allocation~\cite{R075,Mayya2021ResilientTaskAllocation}. Therefore, the communication value of body-state semantics is not only to supplement individual state descriptions, but to transform individual state changes into task-adjustment bases at the collective level. If communication content remains decoupled from these interoceptive semantics, a collective-intelligence framework may still fall into systematic failure because of unsustainable and uncalibrated collaborative task allocation.

\subsection{Environment-Mediated Semantics}
\label{subsec:environment_mediated_semantics}

In conventional wireless-communication frameworks, the physical environment is usually treated as an external condition that affects link quality and sensing reliability. Although some studies have begun to use semantic cues extracted from environmental topology images for predictive millimeter-wave beamforming, such methods still essentially treat environmental features as auxiliary parameters for optimizing low-level physical-layer link configuration~\cite{R013}. In contrast, the conceptual core of ESC indicates that the physical world is no longer only a source of random fading, multipath noise, or line-of-sight blockage. Instead, the environment is elevated at the cognitive level, becoming an active participant in semantic generation and an implicit medium for decentralized collaboration.

Mechanistically, the physical environment itself participates in the basic generation of embodied semantics, rather than merely serving as a static observation background. In different task scenarios, the same environmental cue can be dynamically mapped to different behavioral instructions, directly connecting physical structural entities with the agent's action space~\cite{R078}. For example, in robot-swarm collaboration in an underground disaster area, the environmental semantics of ``slippery ground'' is not simply a description of surface condition. It directly constrains group speed planning, spacing maintenance, and formation adjustment. Therefore, in the ESC framework, the environment can no longer be reduced to a passive object of representation. It becomes a semantic-formation condition that jointly constrains communication content together with body state and task goals.

The environment can also serve as a persistent carrier of operational signals and an asynchronous medium for indirect multi-agent collaboration. Embodied agents can change environmental states through physical actions and leave persistent spatial cues, enabling other agents to obtain action hints when they encounter the same environment. For instance, in swarm UAV search, an early-arriving UAV can write searched regions, suspected target locations, or high-uncertainty regions into a shared map, so that later UAVs can reduce repeated search and adjust inspection directions. This mechanism, in which distributed nodes achieve macro-level synchronized collaboration through local environmental modification, has been rigorously characterized in many related studies~\cite{R084,R085}.

Under the ESC framework, the distinct value of environment-mediated semantics lies in temporal persistence and spatial reusability. It reveals that task states do not need to be resent through explicit messages every time. They can also be stored in the environment or environmental representations and later read by subsequent agents. Therefore, environment-mediated semantics becomes a key system characteristic of ESC: the environment is no longer only the object being described, but can also support semantic formation, task memory, and collaborative cues.

\subsection{Intent Sharing}
\label{subsec:intent_sharing}

In human collaboration, team members with long-term cooperative experience can often achieve efficient synchronization through minimal verbal expressions or simple gestures. Such high communication efficiency does not come from the informational completeness of the isolated signal itself. Rather, the receiver can combine explicit cues with shared experience, environmental semantics, and body-state semantics to decode the sender's intent~\cite{R009}. Extending this cognitive mechanism to collective embodied intelligence, the substantial communication compression brought by long-term repeated collaboration is not simple information omission. Instead, it appears as an intent-sharing mechanism gradually formed among communication nodes through long-term reinforcement and error-feedback correction~\cite{R089}. In ESC systems, \emph{long-term machine collaboration can also form similar tacit coordination}: as task experience accumulates, embodied agents gradually learn which conditions require explicit explanation and which actions can be completed by shared intent and local perception, thereby reducing repeated explanation and explicit signaling transmission.

Unlike conventional data compression, which encodes the same semantic representation into a shorter form, intent sharing emphasizes that multiple embodied agents form stable conditions for intent interpretation through feedback correction and accumulated common experience. Reference~\cite{R088} points out that collaborative communication involves not only information content, but also whether both parties can confirm that they have established the same semantic baseline. Therefore, in the ESC paradigm, the effectiveness of compressed messages does not depend on their structural completeness, but on whether the receiver can use common ground and context-driven inference to complete incomplete and abstract semantics. This nonlinear completion mechanism of receiver $j$ at time $t$ can be formalized as
\begin{equation}
\hat{\phi}_{j,t}
=
\Pi_j\left(
\tilde{\phi}_{t};
\mathcal{G}_{ij,t}, s_{j,t}, e_t, \tau_t
\right).
\end{equation}
\noindent Here, $\tilde{\phi}_{t}$ denotes an incomplete semantic message, $\mathcal{G}_{ij,t}$ denotes the common ground formed by agents $i$ and $j$ through historical interaction, and $s_{j,t}$, $e_t$, and $\tau_t$ denote the receiver state, environmental state, and current task, respectively. The receiver does not mechanically restore the textual content omitted by the sender. Instead, it generates action understanding $\hat{\phi}_{j,t}$ based on shared intent, local perception, and task context.

Intent sharing enables embodied agents in long-term collaboration to form consistent action understanding with less explicit communication, and is an important system characteristic that distinguishes ESC from one-shot semantic compression. \emph{Communication efficiency comes not only from information compression, but also from reduced interaction-confirmation overhead after intent alignment}. However, when task progress, collaboration members, or environmental conditions change, existing compressed semantics may no longer point to the same action meaning. Therefore, embodied agents in ESC systems need the ability to update common ground, identify intent mismatch, and recalibrate, so that implicit collaborative relations can continue to serve collective tasks in dynamic embodied environments.

\begin{figure}[!t]
  \centering
  \includegraphics[width=\columnwidth]{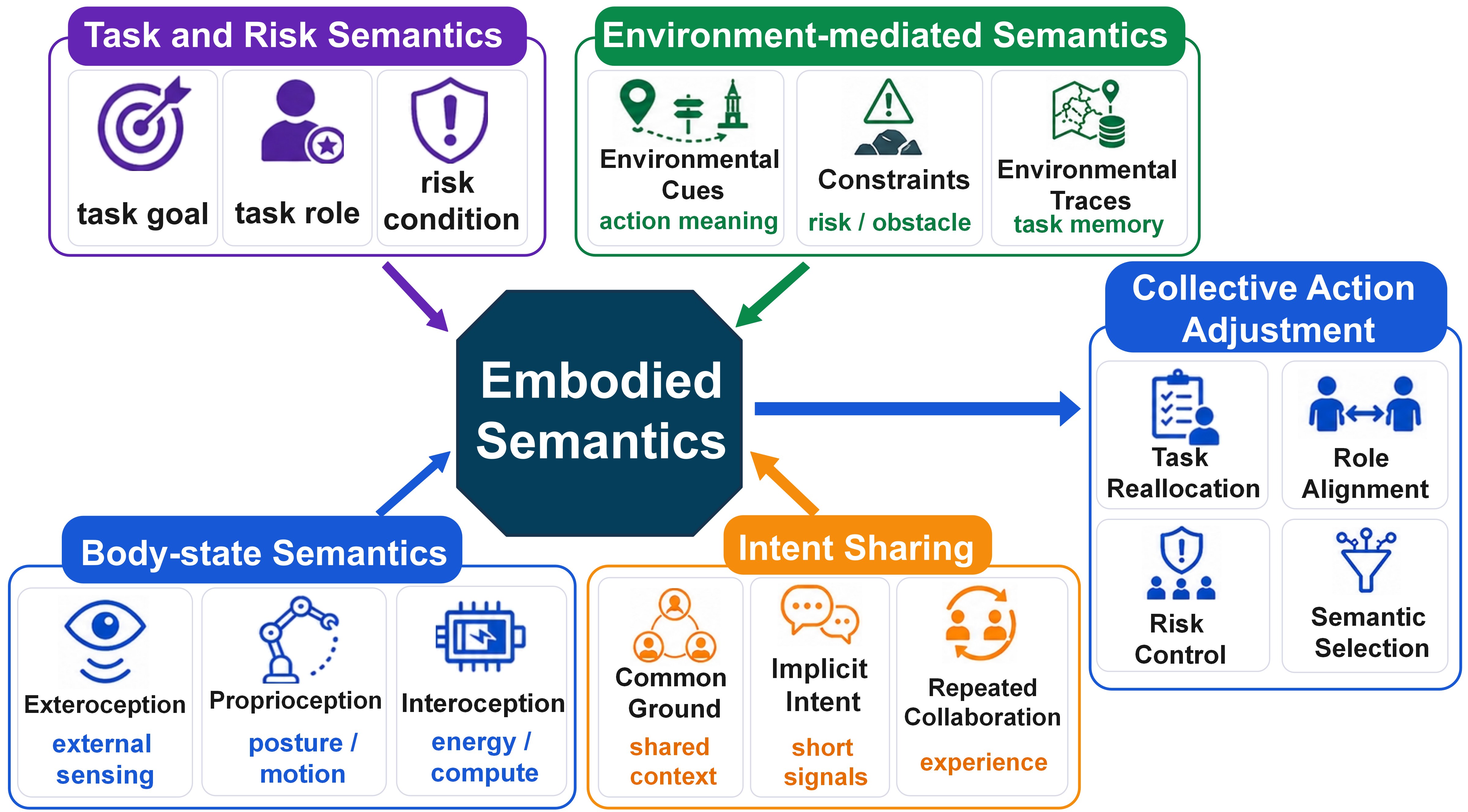}
  \caption{Formation of action-relevant semantics from body state, environment, task goals, and interaction history.}
  \label{fig:action_relevant_semantics}
\end{figure}

Fig.~\ref{fig:action_relevant_semantics} summarizes the formation of action-relevant semantics. Body state, environmental conditions, task goals, and interaction history jointly constrain communication content, allowing semantic selection to directly support collaborative task adjustment.

\subsection{Heterogeneous Embodiment Synchronization}
\label{subsec:heterogeneous_embodiment_synchronization}

In collective embodied collaboration, the system heterogeneity faced by wireless networks extends beyond conventional cross-layer protocol mismatch or hardware-throughput asymmetry. It further appears in the fact that the same semantics can be transformed by heterogeneous agents into different execution modes. For example, UAVs, legged robots, and high-degree-of-freedom manipulators may receive the same task objective, yet generate different local action strategies because their workspaces and execution costs differ. Existing distributed-collaboration studies show that heterogeneous agents can dynamically select collaborative actions according to task requirements and their own capabilities~\cite{Calvo2025LongEnduranceMRTA}. If the communication layer only transmits fixed semantic descriptions while ignoring differences in action mapping across platforms, it remains difficult to guarantee multi-agent action alignment in macro-level collaborative tasks.

The fundamental reason that multi-agent alignment is difficult to achieve seamlessly lies in heterogeneity at multiple levels, including semantics, sensing, and morphology. Specifically, morphological differences determine the executable action spaces and dynamic constraints of different agents; sensing differences restrict individual observation ranges; and differences in semantic reference affect how agents understand the same environmental situation~\cite{R095}. Existing studies show that constructing unified semantic representations can effectively reduce semantic-transmission burden and establish common descriptive foundations for geometric and kinematic parameters~\cite{R094}. However, such unified representations mainly remain in cognitive space. They do not provide the mathematical guarantees needed for real-time motion compatibility, nor can they eliminate local execution mismatches between non-isomorphic mechanical structures.

Therefore, static shared representations can only serve as a foundation. The deeper concern of ESC is how heterogeneous embodied agents continuously align action bases through communication during task execution. Networked multi-agent consensus frameworks have shown that the algebraic topology of the communication network is deeply coupled with the rate of information diffusion and directly affects synchronization in cooperative control~\cite{R097}. Reference~\cite{R098} further points out, from both multi-robot applications and network technologies, that system co-design should include both collaboration algorithms and network systems. Moreover, studies on dynamic task allocation for heterogeneous multi-robot systems show that collective collaboration also requires continuous role and task adjustment according to subtask structure, individual capability differences, and connectivity constraints~\cite{R099,R100,R101}. Together, these studies and arguments point to this core characteristic of ESC: heterogeneous embodiment synchronization is difficult to achieve merely by transmitting more static unified representations. It lies in enabling different agents to continuously transform unified embodied semantics into compatible actions according to their own morphology, action space, and task constraints, and to update them in time during collaborative execution, thereby truly supporting collective embodied tasks.

\section{Environment-Constrained Technical Pathways}
\label{sec:technical_pathways}

Although the above system characteristics require a fundamental reorganization of how communication problems are formulated, ESC systems in the real physical world do not face an idealized semantic transmission process. Dynamically deployed ESC networks are always affected by constraints such as perceptual modality degradation, limited local perception, finite channel resources, and insufficient semantic confidence. Across the entire lifecycle of semantic acquisition, sharing, transmission, and alignment, these multi-level physical constraints introduce severe semantic anomalies.

\begin{figure*}[!t]
  \centering
  \includegraphics[width=0.98\textwidth]{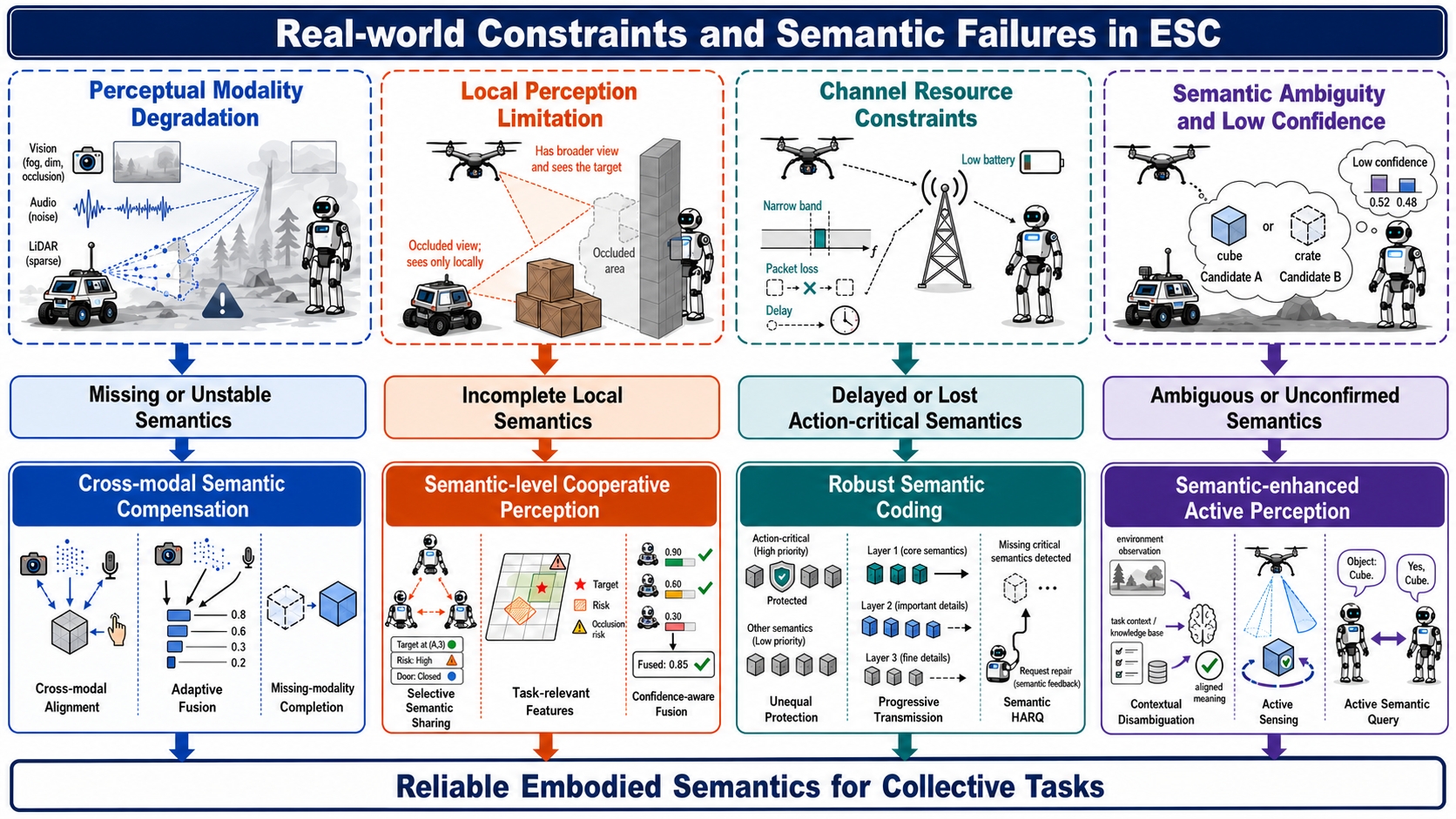}
  \caption{Real-world environmental constraints in embodied semantic communication, their resulting semantic failures, and corresponding technical pathways for ensuring reliable embodied semantics in collective tasks.}
  \label{fig:real_world_constraints}
\end{figure*}

To systematically analyze these failure modes, this section formalizes four types of environmental constraints: missing semantics under perceptual degradation, omitted semantics caused by local observation, transmission risks of critical semantics under channel limitations, and confirmation blockage induced by low confidence. Correspondingly, this paper further proposes directly matched technical responses, answering what actionable semantics should be acquired, when distributed sharing should be performed, how resource-adaptive coding granularity should be optimized, and what confidence threshold is sufficient for closed-loop physical execution. Fig.~\ref{fig:real_world_constraints} systematically organizes the mapping among these constraints, semantic failures, and technical responses.

\subsection{Cross-Modal Semantic Compensation}
\label{subsec:cross_modal_compensation}

In real physical environments, distributed embodied agents are inevitably affected by nonstationary physical-environment dynamics and face multi-level perceptual modality degradation. Random disturbances in raw perceptual links, such as abrupt illumination variations, structural sensor noise, and non-line-of-sight occlusion, can seriously affect the posterior estimation of environmental states and the agent's own proprioceptive configuration~\cite{R105,R106}. At the information level, these physical disturbances appear as instability or even absence of embodied task-related semantics, affecting the operation of embodied task loops.

To mitigate perceptual modality degradation, existing methods mainly start from the degraded modality itself to improve semantic extraction. Recent work has further moved toward scenario-specific semantic enhancement: the work in~\cite{Zhang2026LLSC} proposes LLSC for low-light image semantic communication, coupling semantic extraction under weak illumination with transmission-oriented image reconstruction. FVSF further fuses multi-scale visual cues and enhanced semantic features to strengthen representation learning under limited-sample conditions~\cite{Wang2026FVSF}. Although these methods provide a foundation for semantic acquisition under degraded conditions, fundamental limitations remain when they are extended to the collective and communication-intertwined embodied loops of ESC. The key point is that degraded-modality restoration does not necessarily lead to task-semantic usability. For example, an enhanced image may become clearer in visual quality, yet still fail to support accurate judgment of traversability risk under complex ground conditions or the action impact of occluded targets. When a perceptual modality cannot provide stable information or completely collapses, the system must move from isolated single-modality symbol recovery toward information-theoretic cross-modal semantic compensation~\cite{R016,Sagduyu2024JointSensingSemantic}.

Operationally, cross-modal semantic compensation can be developed at three progressive algorithmic levels: cross-modal alignment, adaptive fusion, and missing-modality semantic completion. At the alignment level, heterogeneous perceptual data streams, including visual sensing information, are projected into a unified and alignable latent tensor space to extract invariant semantic anchors about spatial structure and target features~\cite{R117,R120,R121,R129}. On this basis, the adaptive fusion layer further adjusts the weights of different perceptual information in the fusion process according to modality availability, confidence, and current communication state~\cite{R113,R123,R124,R126}. Finally, at the completion layer, the framework uses unsupervised cross-modal semantic transfer to project prior knowledge representations from structurally more robust sensing channels into degraded modalities, thereby repairing semantic gaps and maintaining task-semantic continuity under severe environmental fluctuations~\cite{R127,R115,R128,R116}.

Together, these pathways show that the core of cross-modal compensation is not simply stacking sensors, but maintaining task-semantic continuity through semantic alignment, reliable fusion, and missing-modality completion. In the ESC framework, this continuity determines whether semantics under degraded conditions can continue to enter collective collaborative tasks and support subsequent communication and action decisions.

\subsection{Semantic-Level Cooperative Perception}

In real environmental interaction, the perceptual field of a single embodied agent is naturally constrained by observation viewpoint, physical occlusion, and finite sensing range. Unlike perceptual modality degradation analyzed in Section~\ref{subsec:cross_modal_compensation}, this spatial limitation does not indicate hardware-level sensor failure, but means that the observable environmental manifold is structurally truncated. ESC architectures therefore need to move from local perception repair toward distributed multi-agent cooperative perception.

However, cooperative perception is not equivalent to direct transmission of raw observations. Raw-data sharing introduces bandwidth burden, and the receiver must still reselect task-relevant content. To address this bandwidth-utility trade-off, cooperative perception under the ESC paradigm focuses on sharing task semantics that support the collective embodied collaboration loop.

The semantic-level cooperative perception framework in ESC can be organized into three interrelated technical sublayers. The first layer shares intermediate semantic representations, compressing raw perception into sparse semantic representations so that receivers obtain environmental information closer to task requirements~\cite{R132,R136,R137,Fu2024ScalableExtraction}. The second layer emphasizes task-relevant communication selection, selecting collaborators~\cite{Xiao2026RFUAVIdentification}, shared regions, and transmission granularity according to the current task to avoid full sharing among all agents~\cite{R133,R138,R139,R140}. The third layer emphasizes confidence-consistent fusion, which screens, corrects, and fuses multi-source shared semantics according to observation viewpoints, semantic confidence, and spatial-overlap relations among different agents, forming collaborative environmental understanding that is more complete and reliable than single-agent observation~\cite{R134,R135,R141,R142}. Under this framework, the local observation of an individual embodied agent can be transformed into communicable and reusable task semantics and enter the collective embodied task loop.

Cooperative perception, however, does not mean that a higher degree of semantic sharing always produces greater system utility. Different embodied agents have different observation conditions, body capabilities, and task roles, and may interpret the same cue differently. A semantic judgment that holds locally for one embodied agent may conflict with the judgments of other agents in the global task~\cite{R144}. Semantics provided by collaborators can also be affected by latency, confidence differences, and context mismatch~\cite{R143,R145}. Therefore, cooperative perception in ESC also needs consistency checking and conflict detection to ensure that the supplemented semantics can truly support stable collective action.

\subsection{Robust Semantic Coding}

Although ESC advances the communication objective from general semantic reproduction to strategic semantic selection and action support, this paradigm shift does not mean that it can escape non-ideal physical wireless channels. In dynamic collective embodied networks, wireless links are always constrained by nonstationary fading, co-channel interference, and limited bandwidth. Resource-allocation frameworks under conventional semantic topologies usually optimize bandwidth allocation only according to semantic reconstruction fidelity~\cite{R103}, whereas ESC faces more stringent operating conditions. What ESC system links transmit is embodied semantics that enters perception, decision-making, and action processes. Therefore, once channel factors cause semantic information to be delayed or lost, even if part of the information eventually arrives, it may have missed the corresponding execution window and can induce unrecoverable collective task failure~\cite{R146,Wang2025AoIResourceAllocation,Xiao2025StatisticalAoI}.

To protect action-critical semantics from non-ideal channel conditions, the basic criterion of network transmission robustness must be fundamentally reconstructed. Although metrics such as bit error rate and packet loss rate form necessary foundations~\cite{R148}, they cannot characterize the heterogeneous task-level consequences caused by different semantic components~\cite{R020}. Existing studies on semantic interference cancellation provide a related foundation for this robustness problem. The works in~\cite{Lin2024SemantIC,Huang2025SwinSemantIC} propose a framework that couples signal-domain processing with semantic-domain reconstruction to suppress wireless interference, and further strengthen this process by improving the representation capability of the semantic recovery model. These works advance interference handling from symbol recovery to semantic recovery, whereas ESC further requires semantic recovery priorities to be determined according to downstream action risk. Based on this robustness requirement under task-risk differences, robust semantic coding in ESC must move beyond a uniform symbol-level protection paradigm and instead develop along three mutually supporting technical pathways: task-oriented unequal error protection, progressive semantic transmission, and semantic-level hybrid automatic repeat request.

At the level of task-oriented unequal error protection, the source bit rate in joint source-channel coding should be dynamically adaptive. It should allocate stronger error-correction redundancy and higher scheduling priority to semantic components according to their mathematically quantified impact on downstream task risk~\cite{R151,R155,R156}. The work in~\cite{Yan2025AdaptiveSemanticGeneration} proposes adaptive semantic generation with NOMA-based interference-aware transmission, providing a concrete example in which semantic components and transmission decisions are jointly adjusted according to channel interference and receiver requirements. Corresponding to this asymmetric protection mechanism, the progressive transmission layer uses hierarchical representation networks to ensure that the minimum usable semantic skeleton arrives deterministically first, and then gradually supplements contextual details when the channel-resource budget allows~\cite{R152,R157,R158}. Semantic successive refinement in~\cite{Zhang2025SemanticSuccessiveRefinement} further realizes this progressive idea by first transmitting coarse semantics and then using generative reconstruction to refine details when resources permit. Finally, at the semantic-level hybrid automatic repeat request layer, retransmission is no longer triggered solely by binary check failure. It is dynamically regulated by semantic-similarity thresholds and task-uncertainty states~\cite{R153,R159,R160}. Only when the receiver's cognitive ambiguity may threaten the stability of the collective embodied loop does the system perform semantic remedial retransmission.

However, the priority of different semantics is itself time-varying and changes with task-stage switching and environmental-state fluctuations. Fixed protection strategies may be mismatched when tasks switch or risks change. Excessive redundancy and retransmission can also occupy resources for other collaborative information and introduce additional latency and scheduling overhead~\cite{Wang2026MultiQoSHRLLC}. Therefore, robust semantic coding in ESC should dynamically adjust protection strategies according to task state and embodied feedback, rather than assigning fixed weights to preset semantic categories.

\subsection{Semantic-Enhanced Active Perception}

Robust semantic coding mainly concerns whether embodied semantics can arrive at the receiver reliably and effectively over constrained channels. Error-free arrival at the syntactic level, however, does not necessarily mean that semantic confidence has been established. Under the ESC paradigm, insufficient semantic confidence mainly appears as cognitive ambiguity, namely the receiver's inability to form the correct mapping for task intent or object reference. Unlike transmission errors in conventional communication that can be located by bit-level detection or block-level checking, such cognitive bias is difficult to eliminate through channel-coding redundancy. If left uncorrected, the receiving agent's internal world state lacks a reliable anchor and further affects downstream perception, decision-making, and collaborative action. Therefore, the receiver must still resolve uncertainty in the transmitted semantic object through autonomous correction, interactive confirmation, or active completion.

The mechanism of semantic-level autonomous correction can be clearly illustrated through contextual understanding in task scenarios. For example, in a warehouse collaboration task, the sender issues the instruction ``bring the tote'', while the receiver may misrecognize it as ``bring the coat'' under noise interference. If the receiver is currently in a packaging area and the surrounding environment mainly consists of logistics boxes, shelves, and packing materials, it can infer from scene observation and task knowledge that the sender is more likely referring to the tote rather than the coat. The correction here does not rely on channel-coding redundancy to recover symbols. It relies on the receiver's semantic alignment of environmental observation, task context, and shared knowledge bases. Shared knowledge bases in semantic communication systems can help transceivers establish interpretable semantic representations and recovery mechanisms, and provide prior support for semantic encoding, transmission, and reconstruction~\cite{R096,Ren2024KnowledgeBaseSemanticCommunication}.

When autonomous ambiguity resolution is insufficient for reliable action decision-making, the receiving agent can further disambiguate through two active mechanisms. The first is \emph{active perception confirmation}, where the agent no longer passively waits for more transmitted information, but actively adjusts observation position, object, or mode according to the current task ambiguity, and aligns newly obtained observations with the existing knowledge base. In this case, perception is no longer only a passive process of receiving external data, but an active information-acquisition process serving task objectives and uncertainty suppression~\cite{Ghasemi2020TaskOrientedActivePerception}. At the same time, new perceptual results still need to be organized into reusable knowledge representations before they can further support subsequent task understanding and execution~\cite{Paulius2019KnowledgeRepresentationRobotics}. The second is \emph{active communication completion}, where the receiver initiates clarification requests to the sender, a human supervisor, or nearby collaborative agents. Through such interactive feedback, the system can further constrain ambiguous task semantics to verifiable action objects or operating conditions, thereby reducing the impact of semantic misunderstanding on subsequent execution~\cite{R183,R184}.

Therefore, active perception in ESC is not a local remedy for a single agent under low-confidence conditions, but \emph{a mechanism for confirming, completing, and aligning transmissible embodied semantics}. Embodied agents can first rely on environmental observation, task context, and shared knowledge bases for semantic-level autonomous correction. If ambiguity may still affect action execution, they further acquire supplementary semantics through active perception or active communication, so that cognitive ambiguity can be handled by the communication system through interaction and correction, thereby supporting the continuous and stable progress of closed-loop action in collective embodied tasks.

\section{Key Theories and Modeling Methods}
\label{sec:modeling_methods}

\begin{figure}[!ht]
  \centering
  \includegraphics[width=\columnwidth]{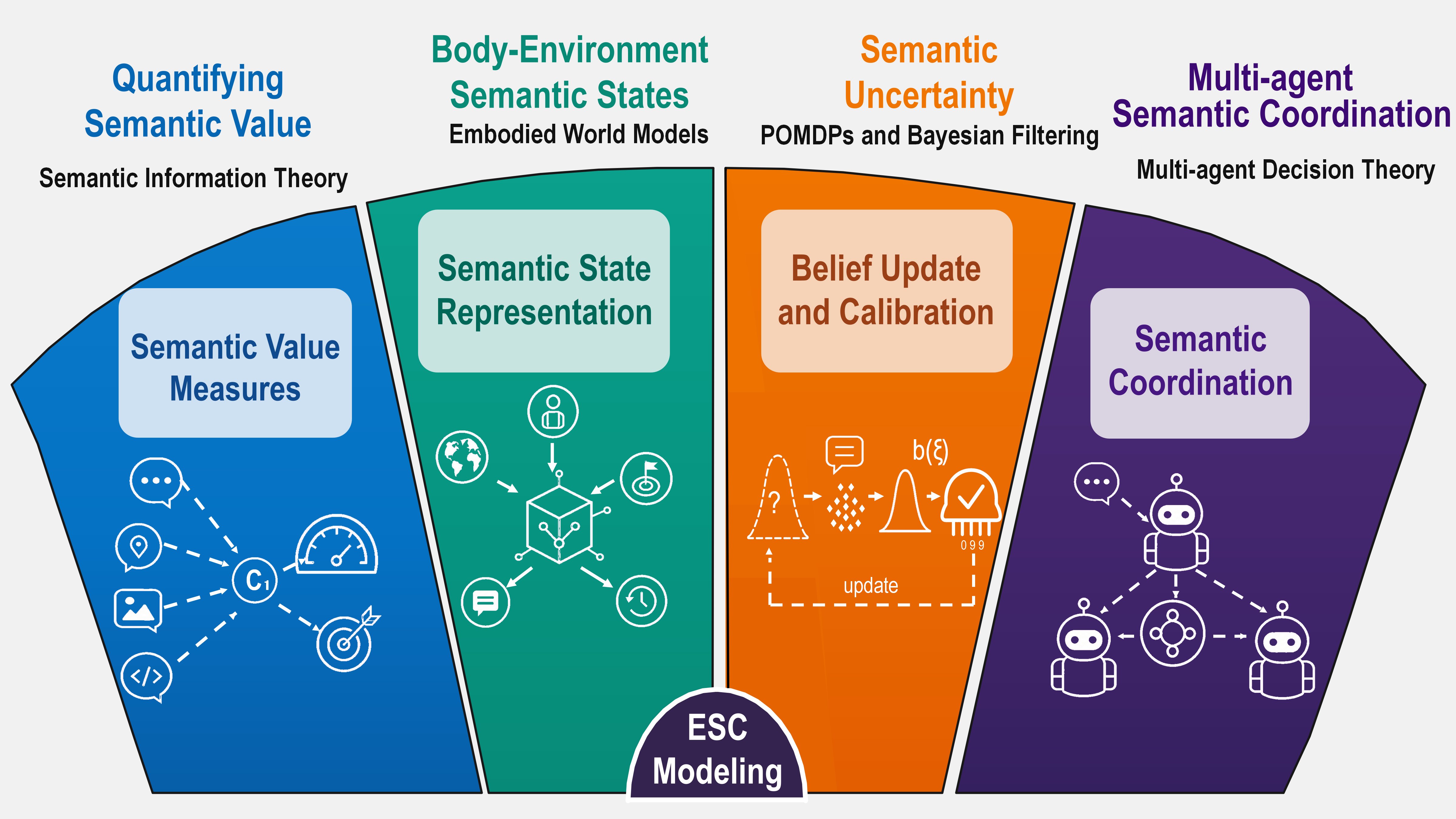}
  \caption{Four modeling questions and their theoretical foundations for ESC.}
  \label{fig:esc_modeling_questions}
\end{figure}

As transmissible embodied semantics become embedded in decentralized collaboration loops, ESC must answer four modeling questions: 1) How should semantic value be quantified? 2) How should semantic state be represented? 3) How should semantic uncertainty be handled? 4) How should multiple agents coordinate their physical actions? Accordingly, this section organizes the relevant foundations into synonymy-based semantic information theory, world models, belief updating and confidence calibration based on partially observable Markov decision processes (POMDPs), and multi-agent decision theory. Fig.~\ref{fig:esc_modeling_questions} summarizes their relationships.

\subsection{Semantic Value Measures}
\label{subsec:semantic_value_measures}

Quantifying semantic value first requires distinguishing semantic information from observable symbols. Niu and Zhang~\cite{NiuZhang2024SemanticTheory,R193} establish a semantic information theory from synonymy, arguing that visible messages such as text, speech, images, and videos belong to syntactic information, while true semantics is implicit in these syntactic forms and must be understood through the receiver's knowledge and context. This view holds that semantics cannot be directly observed like a symbol sequence, and that different syntactic forms can correspond to the same semantics. It provides a clear entry point for semantic-value quantification. Under the ESC paradigm, the criterion for synonymy needs to shift further from abstract semantic similarity to action equivalence in embodied tasks: different environmental descriptions or control signals constitute embodied synonymy only when they guide the receiving agent to form equivalent task judgments, execute the same physical actions, or produce equivalent task outcomes.

In synonymy-based semantic information theory, semantic space $\tilde{U}$ and syntactic space $U$ are connected through a synonymy mapping. If $\tilde{u}_{i_s}$ denotes a semantic element, then the corresponding syntactic synonym set can be written as $U_{i_s}=\{u_i\}_{i\in N_{i_s}}$, where $N_{i_s}$ is the set of all syntactic indices synonymous with this semantic element. This relation can be expressed as
\begin{equation}
f:\tilde{u}_{i_s}\rightarrow U_{i_s}.
\end{equation}
This means that one semantic element can be expressed by multiple syntactic forms, while one syntactic form usually points to a specific semantic element in a given context.

Based on this mapping, semantic communication systems need to introduce a semantic source, a semantic destination, a synonymy mapping, and an inverse synonymy mapping beyond the conventional transmitter, channel, and receiver. The transmitter no longer only encodes symbols, but maps semantics into transmissible syntactic forms. The receiver no longer only recovers symbols, but recovers semantics from the received syntax. This processing logic advances the communication objective from symbol-level consistency to semantic-level consistency. After entering ESC, synonymy mapping must further vary with tasks and body conditions. The same environmental description does not necessarily have the same task meaning for different embodied agents. Therefore, a semantic synonym class cannot be determined only by content similarity. It must also depend on whether it supports identical or similar action judgment.

On the basis of synonymy mapping, semantic entropy can be defined by the probability of the synonym set corresponding to a semantic element. Let $p(U_{i_s})$ denote the probability of semantic element $\tilde{u}_{i_s}$. Then
\begin{equation}
p(U_{i_s})=\sum_{i\in N_{i_s}}p(u_i).
\end{equation}
The semantic entropy is defined as
\begin{equation}
\begin{aligned}
H_s(\tilde{U})
&=-\sum_{i_s=1}^{|\tilde{U}|}p(U_{i_s})\log p(U_{i_s})\\
&=-\sum_{i_s=1}^{|\tilde{U}|}
\left(\sum_{i\in N_{i_s}}p(u_i)\right)
\log\left(\sum_{i\in N_{i_s}}p(u_i)\right).
\end{aligned}
\end{equation}
Because multiple syntactic elements can be merged into the same semantic element, semantic entropy is no greater than classical Shannon entropy:
\begin{equation}
H_s(\tilde{U})\le H(U).
\end{equation}
When each synonym set contains only one syntactic element, semantic entropy degenerates to Shannon entropy. When a synonym set contains multiple syntactic elements, semantic entropy characterizes uncertainty at the meaning level rather than uncertainty at the syntactic-form level.

Synonymy mapping can also extend mutual information, channel capacity, and rate-distortion functions. Reference~\cite{NiuZhang2024SemanticTheory,R193} distinguishes upper semantic mutual information and lower semantic mutual information:
\begin{equation}
I_s^{\uparrow}(\tilde{U};\tilde{V})
=H(U)+H(V)-H_s(\tilde{U},\tilde{V}).
\end{equation}
\begin{equation}
I_s^{\downarrow}(\tilde{U};\tilde{V})
=H_s(\tilde{U})+H_s(\tilde{V})-H(U,V).
\end{equation}
They satisfy the following relation with classical mutual information:
\begin{equation}
I_s^{\downarrow}(\tilde{U};\tilde{V})
\le I(U;V)\le
I_s^{\uparrow}(\tilde{U};\tilde{V}).
\end{equation}
Based on upper semantic mutual information, semantic channel capacity can be defined as
\begin{equation}
C_s=\max_{f_{xy}}\max_{p(x)}I_s^{\uparrow}(\tilde{X};\tilde{Y}).
\end{equation}
Based on lower semantic mutual information, the semantic rate-distortion function can be defined as
\begin{equation}
R_s(D)=
\min_{f_x,f_{\hat{x}}}
\min_{p(\hat{x}|x)\in\mathcal{P}_D}
I_s^{\downarrow}(\tilde{X};\hat{\tilde{X}}).
\end{equation}
Compared with classical capacity and the classical rate-distortion function, their relation can be written as
\begin{equation}
C_s\ge C,\qquad R_s(D)\le R(D).
\end{equation}

These formulas show that semantic communication is not simply reducing the amount of transmitted data. It uses synonym sets to reduce uncertainty and redundancy at the meaning level. As long as the receiver recovers the correct semantics, different syntactic forms under the same semantics do not have to be reproduced one by one.

In summary, synonymy-based semantic information theory provides ESC with a reusable quantitative entry, enabling semantic value to be characterized through synonym sets, semantic entropy, semantic mutual information, and semantic rate-distortion, rather than remaining at raw-bit or symbol-similarity levels. Furthermore, ESC needs to extend content equivalence in general semantic communication to action equivalence.

\subsection{Body--Environment Semantic States}
\label{subsec:body_environment_semantic_states}

Under the ESC framework, before a semantic message enters the transmission, understanding, decision, and action chain, it is necessary to define the state-representation space that it updates inside the robot. In conventional semantic communication, the communication object is usually limited to abstract content semantics. This representation detached from embodied interaction is inherently insufficient for characterizing physical interaction between autonomous entities and dynamic environments. The same embodied semantics may correspond to completely different operational realities under different hardware capabilities, task stages, or historical interaction paths. To address this issue, the ESC paradigm proposes a basic design claim: semantic representation should not be separated from physical reality, but should be modeled as a high-dimensional joint manifold that includes the agent's own body configuration, surrounding environmental topology, long-term task goals, and accumulated interaction history.

World models provide an ideal theoretical entry for this holistic representation. Reference~\cite{Ha2018WorldModels} learns environmental dynamics through compressed spatiotemporal representations, allowing agents to predict action outcomes inside an internal model. Reference~\cite{LeCun2022Path} further proposes configurable predictive world models, emphasizing the representation of perceptual states and action policies at different abstraction levels. Following the structured idea of world models, this paper formalizes the integrated body--environment semantic state of embodied agent $i$ at discrete time $t$ as the following mathematical tuple:

\begin{equation}
\xi_{i,t}
=
\left(
x_{i,t}^{\mathrm{body}},
x_{i,t}^{\mathrm{env}},
g_t,
h_{i,\leq t},
m_{i,\leq t}
\right).
\end{equation}
\noindent where $x_{i,t}^{\mathrm{body}}$ denotes the agent's internal perceptual state, $x_{i,t}^{\mathrm{env}}$ denotes the perceptual state of the external environment, $g_t$ defines the global task goal, $h_{i,\leq t}$ denotes the historical action sequence up to time $t$, and $m_{i,\leq t}$ denotes the sequence of semantic messages received up to time $t$. This formal representation constitutes the mathematical anchor of ESC. It emphasizes that the received communication message $m$ is embedded as a conditional variable in the agent's internal physical world model and further influences the update of its overall state representation.

After defining this complex state vector, the next challenge is to model how this state evolves over time during motion and communication. Because the body--environment semantic state contains self state, external environment, task goal, and interaction history at the same time, its evolution is essentially dynamic coupling and recombination of multi-source information under task constraints. Therefore, embodied world models can serve as an important modeling entry for characterizing this state evolution, and predictive world models are especially suitable for describing structural changes in the representation space. The image-based joint embedding predictive architecture (I-JEPA) minimizes prediction error in the semantic representation domain through the following structured loss function, thereby optimizing latent feature structure~\cite{Assran2023IJEPA}:

\begin{equation}
\mathcal{L}_{\mathrm{I\text{-}JEPA}}
=
\frac{1}{M}
\sum_{k=1}^{M}
\sum_{j\in B_k}
\left\|
\hat{\mathbf{s}}_{y_j}-\mathbf{s}_{y_j}
\right\|_2^2 .
\end{equation}

\noindent Here, $B_k$ denotes the $k$-th target region, and $\hat{\mathbf{s}}_{y_j}$ and $\mathbf{s}_{y_j}$ denote the predicted latent representation and true target feature at position $j$, respectively.

To further characterize the dynamic evolution of states in temporal interaction, recurrent state-space models (RSSMs) model the relation among historical information, actions, and current observations through hidden states, and continuously correct the latent world state using observation feedback. Reference~\cite{Burchi2024MuDreamer} builds the following state-update process based on RSSM:

\begin{equation}
\begin{aligned}
x_t &= \operatorname{enc}_{\phi}(o_t),
&\quad h_t &= f_{\phi}(h_{t-1},z_{t-1},a_{t-1}),\\
z_t &\sim q_{\phi}(z_t\mid h_t,x_t),
&\quad \hat{z}_t &\sim p_{\phi}(\hat{z}_t\mid h_t).
\end{aligned}
\end{equation}

\noindent Here, $x_t$ denotes the observation representation encoded from current observation $o_t$; $h_t$ denotes the recurrent state that fuses historical hidden state, historical latent state, and action input; $z_t$ denotes the posterior latent state after incorporating the current observation, used to correct the agent's estimate of the current world state; and $\hat z_t$ denotes the prior latent state predicted only from historical state. Such models characterize state evolution in dynamic environments through a ``historical prediction--current correction'' mechanism.

On this basis, world models for embodied interaction also need to include action input and historical state in the prediction process, extending semantic-state modeling from static environmental content representation to action-conditioned state transition modeling~\cite{Wu2024iVideoGPT,Zhu2025IRASim,Yang2024STP}.

If semantic state $\xi_{i,t}$ is represented as latent state $z_{i,t}$, the state-prediction process in ESC can be summarized as
\begin{equation}
p_\theta
\left(
z_{i,t+1}
\mid
h_{i,t},
z_{i,t},
a_{i,t},
m_{i,t}^{\mathrm{sem}}
\right).
\end{equation}
\noindent where $h_{i,t}$ denotes historical state information, $a_{i,t}$ denotes the action of the agent, and $m_{i,t}^{\text{sem}}$ denotes semantic information received from the channel. This formula constitutes the core operation mechanism of world models in ESC: it explicitly introduces communication semantic messages as key conditioning variables for correcting the receiver's prediction of physical-environment evolution and self state.

In addition, when data-driven state-transition models face out-of-distribution scenarios, knowledge-driven methods can further integrate explicit physical priors, commonsense knowledge bases, and task-boundary rules into transition-probability models~\cite{Lim2022Real2Sim2Real,Wong2023WordToWorld}. Ultimately, world models provide ESC with a unified entry for semantic-state modeling, allowing self state, environmental state, task goals, interaction history, and communication messages to be incorporated into a unified representation.

\subsection{Semantic Uncertainty}
\label{subsec:semantic_uncertainty}

Even with relatively complete predictive world models, the receiver still needs to evaluate the reliability of its semantic understanding to avoid incorrect semantic judgments under noise interference and model bias. In real physical environments, high-dimensional joint states are usually not directly observable. Agents can only perform probabilistic estimation based on limited sensing information and model inference, and this process inevitably involves uncertainty. According to the classic classification in deep learning, uncertainty mainly comes from two sources: aleatoric uncertainty caused by sensing noise and data ambiguity, and epistemic uncertainty caused by insufficient model knowledge and environmental distribution shift~\cite{R165}. Mapped to ESC systems, observation incompleteness mainly increases aleatoric uncertainty, while insufficient model generalization further amplifies epistemic uncertainty. Therefore, actionable semantic representations in ESC should provide not only a single state estimate, but also the reliability of that estimate.

Partially observable Markov decision processes (POMDPs) provide a basic modeling framework for this probabilistic state representation. To mathematically formalize this semantic uncertainty state, ESC models the agent's semantic understanding as an updatable probability distribution over the state space. Based on the classical POMDP modeling framework~\cite{R231}, the internal belief state $b_{i,t}$ denotes the agent's posterior probability distribution over physical-world states that cannot be directly observed, and this distribution is recursively constructed through historical trajectory information and active observation processes~\cite{R232}. After a formal extension to the ESC system architecture, the belief distribution $b_{i,t}(\xi)$ maintained by embodied agent $i$ at time $t$ over its integrated body--environment semantic state $\xi_{i,t}$ can be defined as

\begin{equation}
b_{i,t}(\xi)
=
\Pr\!\left(
\xi_{i,t}=\xi
\mid
h_{i,\leq t},m_{i,\leq t}
\right).
\end{equation}

This equation shows that ESC incorporates received semantic messages into state-belief updating, enabling the receiver to represent potential physical states as a probability distribution rather than relying on a single deterministic estimate.

After semantic representation is modeled as a probability distribution, it is further necessary to characterize how newly received semantic information participates in recursive semantic uncertainty updating and gradually corrects the agent's belief estimate of the current state. Conventional communication systems mainly focus on accurate recovery of message content. In the ESC loop, however, the received semantic message needs to be incorporated as additional information into the belief-update process, together with historical information, to reduce the receiver's uncertainty about the body--environment semantic state. Following the prediction-correction mechanism in the POMDP framework, the information-driven semantic uncertainty update can be formalized as
\begin{equation}
\begin{aligned}
\bar b_{i,t}(\xi)
&=
\int_{\Xi}
p\!\left(
\xi\mid\xi',a_{i,t-1}
\right)
b_{i,t-1}(\xi')\,\mathrm{d}\xi',\\
b_{i,t}(\xi)
&=
\eta\,
p\!\left(
m_{i,t}^{\mathrm{sem}}
\mid \xi,c_{i,t}
\right)
\bar b_{i,t}(\xi).
\end{aligned}
\end{equation}
\noindent where $\bar b_{i,t}(\xi)$ denotes the prior predictive distribution obtained from the state-transition model and the belief distribution at the previous time, $\eta$ is a normalization factor, and $c_{i,t}$ denotes external conditions affecting the reliability of semantic information, including communication link state. This recursion can be viewed as an extension of classical Bayesian filtering in ESC scenarios. The received semantic message $m_{i,t}^{\mathrm{sem}}$ participates in posterior belief updating through a conditional likelihood model, enabling the receiver to dynamically correct the probability estimate of the body--environment semantic state according to communication reliability, thereby continuously quantifying and reducing semantic uncertainty.

However, representing the semantic state as a probability distribution alone is not sufficient to guarantee its reliability. The key is whether this probability estimate truly reflects the trustworthiness of actual state inference. Existing studies show that deep neural networks commonly suffer from overconfidence, meaning that high model confidence does not necessarily correspond to higher actual prediction accuracy. Therefore, in ESC systems, semantic-state inference needs not only to characterize uncertainty, but also to calibrate confidence estimates.

Let $\hat{\xi}_{i,t}$ denote the maximum a posteriori state estimate obtained from the current belief distribution:

\begin{equation}
\hat{\xi}_{i,t}
=
\arg\max_{\xi} b_{i,t}(\xi).
\end{equation}

Further define a confidence function $C_{i,t}=g(b_{i,t})$ derived from the belief distribution. Under ideal calibration, it should satisfy

\begin{equation}
\Pr
\left(
\xi_{i,t}=\hat{\xi}_{i,t}
\mid
C_{i,t}=q
\right)
\approx q .
\end{equation}

This relation indicates that when the model assigns confidence $q$ to a semantic-state estimate, its actual long-term statistical correctness rate should be consistent with that confidence. For ESC, this calibration mechanism can prevent unreliable semantic states from being directly used for physical decision-making, and provides a reliable basis for subsequent risk-aware control, conservative execution, or active semantic acquisition.

\subsection{Multi-Agent Semantic Coordination}
\label{subsec:multi_agent_semantic_coordination}

The preceding discussion mainly concerns how a single embodied agent quantifies semantic value, estimates semantic state, and handles semantic uncertainty. In the collective embodied-intelligence collaboration scenarios considered by ESC, however, multiple agents need to form coordinated actions based on limited local observations and communication information. The core problem further becomes how to evaluate the contribution of semantic messages to collective tasks and effectively integrate them into joint decision-making. Therefore, the modeling framework of ESC needs to extend from single-agent state estimation to distributed multi-agent collaborative decision-making. Markov games and POMDPs in multi-agent reinforcement learning (MARL) provide basic mathematical frameworks for describing local observation, communication interaction, and joint action~\cite{Zhang2026MAHEADNet}.

In collective collaborative tasks, agents usually have only local observations, while the behaviors of different agents jointly affect environmental evolution and overall task benefit. Reference~\cite{Lowe2017MADDPG} uses a partially observable Markov game to model this process, in which multiple agents select policies based on local information and the coupling among different agents' behaviors affects subsequent state changes and task outcomes. Under local information constraints, the decision policy of embodied agent $i$ can be written as

\begin{equation}
\pi_{\theta_i}(a_i|o_i).
\end{equation}
\noindent where $o_i$ denotes the local observation obtained by agent $i$, and $a_i$ denotes the action it selects.

The joint actions of multiple agents affect subsequent state evolution through environmental dynamics as:

\begin{equation}
T:\mathcal{S}\times\mathcal{A}_1\times\cdots\times\mathcal{A}_N\rightarrow\mathcal{S}.
\end{equation}

\noindent Here, $\mathcal{S}$ denotes the global environmental state space, $\mathcal{A}_i$ denotes the action space of the $i$-th embodied agent, and $T$ denotes the environmental state-evolution function driven by joint actions. This modeling unifies local decision-making of individual embodied agents and collective collaborative behavior in the same framework.

A further key issue in multi-agent collaboration is how to learn effective policies with limited information. The centralized training with decentralized execution (CTDE) paradigm allows the training stage to use global states and other agents' action information to estimate joint value, while the execution stage still maintains independent decision-making based on local observations and communication information. Based on this idea, multi-agent actor--critic methods evaluate joint action value through a centralized critic, and their policy gradient can be expressed as

\begin{equation}
\nabla_{\theta_i}J(\theta_i)
=
\mathbb{E}
\left[
\nabla_{\theta_i}\mu_i(o_i)
\nabla_{a_i}
Q_i(x,a_1,\ldots,a_N)
\right].
\end{equation}

\noindent Here, $\theta_i$ denotes the policy parameter of agent $i$, $J(\theta_i)$ denotes the expected accumulated return of the corresponding policy, and $o_i$ and $a_i$ denote the local observation and action of this agent, respectively. $Q_i(x,a_1,\ldots,a_N)$ denotes the centralized critic's estimate of the current policy value based on global state $x$ and all agents' joint actions during training, while $\nabla_{a_i}Q_i$ characterizes how a change in agent $i$'s action affects joint task benefit. This method uses global information to learn collaborative policies during training, while keeping decentralized decision-making based on local observations during execution, providing a modeling foundation for using semantic information to assist multi-agent collaboration in ESC.

Decentralized execution, however, does not mean that local policies naturally form globally optimal collaboration. Because different agents may produce conflicting behaviors based on local information, it is necessary to further characterize the consistency relation between individual value and collective value. The value-decomposition method QMIX represents the joint action value as a monotonic combination of individual value functions, so that global value maximization remains consistent with local action selection~\cite{Rashid2018QMIX}:

\begin{equation}
\frac{\partial Q_{\mathrm{tot}}}{\partial Q_a}\geq0,\quad \forall a .
\end{equation}

\noindent Here, $Q_{\mathrm{tot}}$ denotes the global value function corresponding to the joint actions of all embodied agents, and $Q_a$ denotes the local action value of the $a$-th embodied agent. This monotonic constraint guarantees consistency between joint value evaluation and individual value selection, so that collaborative behavior optimized by global task benefit can be implemented through each agent's local decision. For ESC, the role of semantic messages must ultimately be reflected in collaborative action decisions, namely whether communication information helps agents evaluate candidate actions more accurately and improve collective task benefit.

On this basis, multi-agent ESC also needs to further address how semantic information should be efficiently exchanged, namely dynamically determining communication targets and communication timing. Fully connected broadcasting can guarantee information coverage, but easily causes redundant communication in resource-constrained environments. Attention-based communication mechanisms learn the matching relation between sender information and receiver needs, allowing communication topology to adjust dynamically with task state. For example, TarMAC represents the communication process as attention matching between sender message representations and receiver query vectors~\cite{R244}:

\begin{equation}
\begin{aligned}
m_i^t &= [k_i^t,v_i^t],\\
\boldsymbol{\alpha}_{j}
&=
\operatorname{softmax}
\left(
\frac{{q_j^{t+1}}^T k_i^t}{\sqrt{d_k}}
\right),\\
c_j^{t+1}
&=
\sum_{i=1}^{N}\alpha_{ji}v_i^t .
\end{aligned}
\end{equation}

\noindent Here, $k_i^t$ denotes the key representation used by the sending agent to describe information features, $v_i^t$ denotes the actual message content, $q_j^{t+1}$ denotes the query representation generated by the receiving embodied agent according to the current task state, and $\alpha_{ji}$ characterizes how much attention the receiver pays to different sending information. This mechanism enables agents to select more relevant semantic information for interaction according to current task needs, reducing redundant communication. It is consistent with the collective task-oriented nature of ESC, making the communication process focus more on semantic information that has practical influence on state understanding and collaborative action.
In summary, multi-agent coordination theory provides ESC with a unified modeling framework from local semantic cognition to collective action decision-making, allowing semantic messages, agent states, individual actions, and collective task benefits to be analyzed within the same system.

\section{Illustrative Case Study: Heterogeneous-Agent Collaborative Tracking}
\label{sec:case_study}

\subsection{Scenario and Conventional Communication Schemes}

\begin{figure*}[!t]
  \centering
  \includegraphics[width=0.98\textwidth]{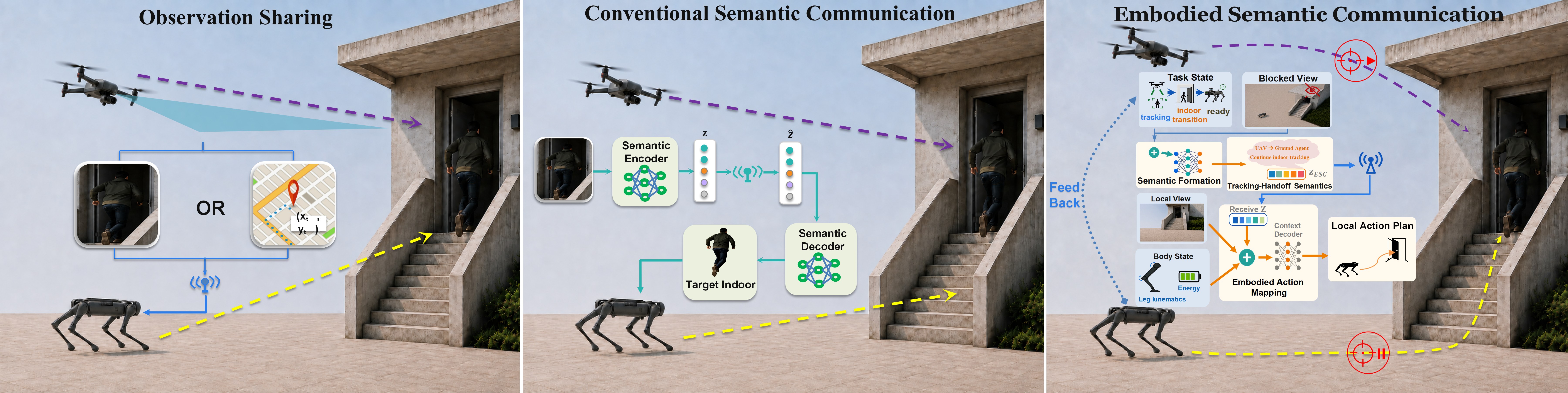}
  \caption{Illustrative comparison of three communication schemes for heterogeneous-agent collaborative tracking. Observation sharing transmits a target image or coordinate update; conventional semantic communication compresses the observation into a latent representation and reconstructs the target semantics at the receiver; ESC updates embodied semantics according to task progress and agent states, forms tracking-handoff semantics when the UAV's view becomes blocked, and enables the quadruped to map the message into a tracking action using its local perception and dynamical capabilities. Execution feedback then updates the collective task state.}
  \label{fig:esc_handoff_example}
\end{figure*}

As shown in Fig.~\ref{fig:esc_handoff_example}, consider a collaborative target-tracking task involving a UAV and a quadruped robot. The UAV follows a moving target in an open area, but the target subsequently enters an enclosed indoor space, requiring the system to adapt its collaboration as environmental conditions change. The figure compares observation sharing, conventional semantic communication, and ESC under this environmental transition.

Under observation sharing, the UAV transmits either a target image or a coordinate update. The image preserves rich identity cues but consumes substantial bandwidth and requires the quadruped to match the target across different viewpoints. The coordinate is compact, but it must be continuously refreshed as the target moves. Conventional semantic communication instead encodes the UAV observation into a compact latent representation and reconstructs target semantics at the quadruped, reducing the payload while retaining information such as target location and motion direction. As long as the UAV continuously observes the target, these schemes can maintain target awareness across the two agents.

Once the target enters the building, however, the UAV's observation becomes occluded, while spatial restrictions and collision risks prevent it from safely continuing the pursuit. Observation sharing can no longer provide a valid image, and the latest coordinate rapidly becomes stale. Conventional semantic communication can compress an available observation but cannot recover a target state that is no longer observed. Ordinary task-oriented communication could partly address this limitation by optimizing a compact message or direct handoff decision for the predefined tracking objective. However, deciding that tracking should be transferred does not determine how a heterogeneous receiver should execute the handoff under its own physical state and local environment.

\subsection{Beyond Task-Specific Handoff Decisions}

In ordinary task-oriented communication, the semantic representation is typically optimized around \emph{predefined} task outputs, receiver models, and action spaces. In this case, it may indicate that the quadruped should take over tracking. Yet the UAV cannot fully determine whether the quadruped has sufficient indoor visibility, mobility, energy, or confidence to execute that decision. When these conditions differ from the predefined receiver model, a fixed task output does not preserve a reusable semantic basis for receiver-dependent action grounding.

ESC instead communicates an embodied semantic state rather than only a handoff result. Occlusion changes semantic value: target position and motion cues lose utility, whereas target identity cues, blocked-view status, task progress, and handoff intent become action-critical. The UAV therefore forms tracking-handoff semantics from these elements. Based on common ground and interaction history, the quadruped interprets the message as a change in collaborative intent, fuses it with its local observation and body state, and maps the collective tracking objective to a platform-specific action, such as entering the building, changing its search region, or requesting clarification. This process instantiates intent sharing, semantic-level cooperative perception, and heterogeneous embodiment synchronization.

At the same time, unlike ordinary task-oriented communication, ESC uses execution feedback not merely to report task success or failure, but also to update the shared embodied semantic state. Target reacquisition, persistent ambiguity, or failed indoor access may trigger semantic supplementation, action adjustment, or role reassignment. Communication therefore evolves with physical execution rather than terminating after a predefined task output is delivered. This forms a closed loop of semantic formation, transmission, action mapping, and feedback-driven updating.

In this case, observation sharing delivers target evidence, conventional semantic communication reconstructs target meaning, and ordinary task-oriented communication may optimize a predefined handoff decision. ESC further enables that intent to be grounded across heterogeneous embodiments and continually updated through physical execution feedback. Its effectiveness should therefore be assessed through tracking continuity, target-reacquisition delay, and execution safety. Receiver-dependent action grounding and feedback-driven role reconfiguration constitute the additional capabilities that ESC provides beyond ordinary task-oriented communication.

\section{Open Challenges and Future Directions}
\label{sec:open_challenges}

Although ESC provides a new communication paradigm for multi-agent collaboration, its practical deployment still faces key theoretical and engineering challenges. Conventional semantic communication mainly evaluates static information in controlled networks. New bottlenecks emerge when semantic links enter dynamic physical action processes and support heterogeneous-agent collaboration. This section systematically summarizes the following major open challenges faced by ESC and further outlines future research directions.

\subsection{Measurable Semantic Reliability Under Dynamic Environments}

Semantic reliability in real ESC systems cannot be predicted accurately by a single model. Changes in the environment, link conditions, and task progress affect semantic usability and may alter its action consequences. A more feasible direction is therefore to develop semantic reliability mechanisms that are measurable, calibratable, and updatable online, rather than pursue fully precise end-to-end uncertainty analysis. Future work should determine how to estimate semantic reliability, when semantic reliability requires recalibration, and how reliability should guide reception, clarification, or degradation decisions.

\subsection{Ambiguity-Triggered Semantic Interaction Under Dynamic Tasks}

If semantic consistency in dynamic tasks remains only a matter of understanding and alignment, it becomes primarily a perception or cognition problem. ESC must instead turn semantic ambiguity into a communication-interaction decision. Embodied agents first interpret messages using local observations, shared context, and knowledge bases; limited interaction is triggered only when ambiguity or misunderstanding may affect action. Future work should identify ambiguity that warrants interaction and determine the content and timing of clarification, so that communication supports continual alignment during task execution.

\subsection{Bandwidth-Adaptive Unified Semantics for Heterogeneous Embodiment}

The communication challenge in heterogeneous embodied collaboration is to transform multimodal observations into unified embodied semantics that can be shared, reused, and parsed on demand by different receivers. Senders must select semantic granularity and necessary redundancy according to link conditions, receiver needs, and ambiguity risk, thereby reducing ineffective transmission. Future work should develop bandwidth-adaptive unified-semantic generation and selective communication mechanisms that distinguish information requiring explicit transmission from information receivers can complete using their own states and local environments.

\subsection{Safety and Trustworthiness in ESC}

Once semantic errors enter embodied action loops, they can turn understanding deviations into physical risks. Unreliable completion, unresolved ambiguity, and attacks on semantic links may all induce incorrect actions. ESC safety is particularly challenging because receivers can immediately update state estimates and action strategies from semantic messages. Future systems should couple trustworthiness assessment with action constraints, and trigger pausing, degradation, verification, or human intervention when semantics is untrustworthy.

\subsection{Task-Level Evaluation Frameworks for ESC}

Establishing comparable evaluation frameworks remains essential for ESC. Link quality, semantic similarity, and task completion each capture only part of communication, expression, or execution, and cannot reveal the practical contribution of semantic communication to collective action. ESC evaluation should connect communication contributions to semantic understanding and task actions, while using controlled comparisons to distinguish the effects of communication, semantic representation, and execution. Reproducible benchmarks across scenarios, agent types, and task objectives are necessary for ESC to progress from a conceptual framework to a verifiable engineering system.

\section{Conclusion}

This paper proposes ESC for collective embodied intelligence as a new communication paradigm. ESC extends the communication object from conventional general semantic representations to dynamic embodied semantics, enabling heterogeneous agents to efficiently parse and align them and further translate them into executable action bases. Under this framework, communication effectiveness is no longer measured solely by conventional bit reliability or semantic reconstruction fidelity, but is grounded in collective closed-loop action outcomes in real physical environments. To systematically establish this paradigm, this paper characterizes ESC from five dimensions, including action-oriented effectiveness. In response to physical constraints in real-world environments, it further discusses key technical pathways such as cross-modal semantic compensation. In addition, the paper examines the mathematical foundations required for ESC modeling and incorporates theories such as synonymy-based semantic value measurement into a unified modeling framework. Looking forward, ESC requires further progress in semantic reliability measurement under dynamic environments and ambiguity-triggered semantic interaction. Overall, ESC shifts the focus of communication research from information transmission and semantic reproduction toward supporting collective action with embodied semantics, providing a new communication perspective for collective embodied perception and collaboration in real physical environments.

\addtolength{\voffset}{-0.07in}
\addtolength{\textheight}{0.14in}
\vspace*{-0.25\baselineskip}
\hypersetup{urlcolor=black}
\bibliographystyle{IEEEtran}
\bibliography{references}

@IEEEtranBSTCTL{IEEEtranBSTCTLNoDash,
  CTLdash_repeated_names = {no}
}

@article{R001,
  author       = {Shannon, Claude E.},
  title        = {A Mathematical Theory of Communication},
  journal      = {Bell Syst. Tech. J.},
  volume       = {27},
  number       = {3},
  pages        = {379--423},
  year         = {1948},
  month        = jul,
  doi          = {10.1002/j.1538-7305.1948.tb01338.x}
}

@article{R002,
  author       = {Weaver, Warren},
  title        = {Recent Contributions to the Mathematical Theory of Communication},
  journal      = {ETC: A Review of General Semantics},
  volume       = {10},
  number       = {4},
  pages        = {261--281},
  year         = {1953},
  month        = {Summer}
}

@article{R004,
  author       = {Qin, Zhijin and Liang, Le and Wang, Zijing and Jin, Shi and Tao, Xiaoming and Tong, Wen and Li, Geoffrey Ye},
  title        = {{AI} Empowered Wireless Communications: From Bits to Semantics},
  journal      = {Proc. IEEE},
  volume       = {112},
  number       = {7},
  pages        = {621--652},
  year         = {2024},
  month        = jul,
  doi          = {10.1109/jproc.2024.3437730}
}

@article{R005,
  author       = {Niu, Kai and Dai, Jincheng and Yao, Shengshi and Wang, Sixian and Si, Zhongwei and Qin, Xiaoqi and Zhang, Ping},
  title        = {A Paradigm Shift toward Semantic Communications},
  journal      = {IEEE Commun. Mag.},
  volume       = {60},
  number       = {11},
  pages        = {113--119},
  year         = {2022},
  month        = nov,
  doi          = {10.1109/mcom.001.2200099}
}

@article{R003,
  author       = {Calvanese Strinati, Emilio and Barbarossa, Sergio},
  title        = {{6G} networks: Beyond {Shannon} towards semantic and goal-oriented communications},
  journal      = {Comput. Netw.},
  volume       = {190},
  year         = {2021},
  month        = may,
  note         = {{Art. no. 107930}},
  doi          = {10.1016/j.comnet.2021.107930}
}

@article{R006,
  author       = {Getu, Tilahun M. and Kaddoum, Georges and Bennis, Mehdi},
  title        = {A Survey on Goal-Oriented Semantic Communication: Techniques, Challenges, and Future Directions},
  journal      = {IEEE Access},
  volume       = {12},
  pages        = {51223--51274},
  year         = {2024},
  month        = mar,
  doi          = {10.1109/access.2024.3381967}
}

@article{Sun2024EmbodiedIntelligenceSurvey,
  author       = {Sun, Fuchun and Chen, Runfa and Ji, Tianying and Luo, Yu and Zhou, Huaidong and Liu, Huaping},
  title        = {A Comprehensive Survey on Embodied Intelligence: Advancements, Challenges, and Future Perspectives},
  journal      = {CAAI Artificial Intelligence Research},
  volume       = {3},
  year         = {2024},
  month        = dec,
  note         = {{Art. no. 9150042}},
  doi          = {10.26599/air.2024.9150042}
}

@article{MonWilliams2025EmbodiedLLMRobot,
  author       = {Mon-Williams, Ruaridh and Li, Gen and Long, Ran and Du, Wenqian and Lucas, Christopher G.},
  title        = {Embodied large language models enable robots to complete complex tasks in unpredictable environments},
  journal      = {Nat. Mach. Intell.},
  volume       = {7},
  number       = {4},
  pages        = {592--601},
  year         = {2025},
  month        = apr,
  doi          = {10.1038/s42256-025-01005-x}
}

@article{Petersen2019CollectiveRoboticConstruction,
  author       = {Petersen, Kirstin H. and Napp, Nils and Stuart-Smith, Robert and Rus, Daniela and Kovac, Mirko},
  title        = {A review of collective robotic construction},
  journal      = {Sci. Robot.},
  volume       = {4},
  number       = {28},
  year         = {2019},
  month        = mar,
  note         = {{Art. no. eaau8479}},
  doi          = {10.1126/scirobotics.aau8479}
}

@article{Talamali2021WhenLess,
  author       = {Talamali, Mohamed S. and Saha, Arindam and Marshall, James A. R. and Reina, Andreagiovanni},
  title        = {When less is more: Robot swarms adapt better to changes with constrained communication},
  journal      = {Sci. Robot.},
  volume       = {6},
  number       = {56},
  year         = {2021},
  month        = jul,
  note         = {{Art. no. eabf1416}},
  doi          = {10.1126/scirobotics.abf1416}
}

@article{Zhu2024CommMADRL,
  author       = {Zhu, Changxi and Dastani, Mehdi and Wang, Shihan},
  title        = {A survey of multi-agent deep reinforcement learning with communication},
  journal      = {Auton. Agents Multi-Agent Syst.},
  volume       = {38},
  number       = {1},
  year         = {2024},
  month        = jan,
  note         = {{Art. no. 4}},
  doi          = {10.1007/s10458-023-09633-6}
}

@article{Chen2026TaskDrivenSemanticCollaborativeCommunication,
  author       = {Chen, Mingkai and Zeng, Mujian and Ma, Wenbo and He, Xiaoming and Al-Dulaimi, Anwer and Mumtaz, Shahid},
  title        = {Task-Driven Semantic Collaborative Communication Helps Multi-Robot Systems with Embodied Intelligence},
  journal      = {IEEE Commun. Mag.},
  volume       = {64},
  number       = {4},
  pages        = {42--49},
  year         = {2026},
  month        = apr,
  doi          = {10.1109/mcom.001.2500506}
}

@article{R028,
  author       = {Malik, Sumbal and Khan, Muhammad Jalal and Khan, Manzoor Ahmed and El-Sayed, Hesham},
  title        = {Collaborative Perception---The Missing Piece in Realizing Fully Autonomous Driving},
  journal      = {Sensors},
  volume       = {23},
  number       = {18},
  year         = {2023},
  month        = sep,
  note         = {{Art. no. 7854}},
  doi          = {10.3390/s23187854}
}

@article{R029,
  author       = {Cui, Guangzhen and Zhang, Weili and Xiao, Yanqiu and Yao, Lei and Fang, Zhanpeng},
  title        = {Cooperative Perception Technology of Autonomous Driving in the Internet of Vehicles Environment: A Review},
  journal      = {Sensors},
  volume       = {22},
  number       = {15},
  year         = {2022},
  month        = jul,
  note         = {{Art. no. 5535}},
  doi          = {10.3390/s22155535}
}

@inproceedings{R030,
  author       = {Jia, Yukuan and Mao, Ruiqing and Sun, Yuxuan and Zhou, Sheng and Niu, Zhisheng},
  title        = {Online {V2X} Scheduling for Raw-Level Cooperative Perception},
  booktitle    = {Proc. IEEE Int. Conf. Commun. (ICC)},
  address      = {Seoul, Republic of Korea},
  pages        = {309--314},
  year         = {2022},
  month        = may,
  doi          = {10.1109/icc45855.2022.9838439}
}

@article{R031,
  author       = {Lin, Chunmian and Tian, Daxin and Duan, Xuting and Zhou, Jianshan and Zhao, Dezong and Cao, Dongpu},
  title        = {{V2VFormer}: Vehicle-to-Vehicle Cooperative Perception With Spatial-Channel Transformer},
  journal      = {IEEE Trans. Intell. Veh.},
  volume       = {9},
  number       = {2},
  pages        = {3384--3395},
  year         = {2024},
  month        = feb,
  doi          = {10.1109/tiv.2024.3353254}
}

@article{R064,
  author       = {Liu, Chenguang and Chen, Yunfei and Chen, Jianjun and Payton, Ryan and Riley, Michael and Yang, Shuang-Hua},
  title        = {Cooperative Perception With Learning-Based {V2V} Communications},
  journal      = {IEEE Wireless Commun. Lett.},
  volume       = {12},
  number       = {11},
  pages        = {1831--1835},
  year         = {2023},
  month        = nov,
  doi          = {10.1109/lwc.2023.3295612}
}

@misc{R039,
  author        = {Qin, Zhijin and Tao, Xiaoming and Lu, Jianhua and Tong, Wen and Li, Geoffrey Ye},
  title         = {Semantic Communications: Principles and Challenges},
  year          = {2022},
  month         = jan,
  howpublished  = {arXiv:2201.01389},
  doi           = {10.48550/arxiv.2201.01389},
  eprint        = {2201.01389},
  archivePrefix = {arXiv},
  primaryClass  = {cs.IT},
  url           = {https://arxiv.org/abs/2201.01389}
}

@article{R040,
  author       = {Zhang, Ping and Liu, Yiming and Song, Yile and Zhang, Jiaxiang},
  title        = {Advances and Challenges in Semantic Communications: A Systematic Review},
  journal      = {Natl. Sci. Open},
  volume       = {3},
  number       = {4},
  year         = {2024},
  month        = jul,
  note         = {{Art. no. 20230029}},
  doi          = {10.1360/nso/20230029}
}

@article{R066,
  author       = {Goldreich, Oded and Juba, Brendan and Sudan, Madhu},
  title        = {A theory of goal-oriented communication},
  journal      = {J. ACM},
  volume       = {59},
  number       = {2},
  pages        = {8:1--8:65},
  year         = {2012},
  month        = apr,
  doi          = {10.1145/2160158.2160161}
}

@article{R007,
  author       = {Getu, Tilahun M. and Kaddoum, Georges and Bennis, Mehdi},
  title        = {Making Sense of Meaning: A Survey on Metrics for Semantic and Goal-Oriented Communication},
  journal      = {IEEE Access},
  volume       = {11},
  pages        = {45456--45492},
  year         = {2023},
  month        = apr,
  doi          = {10.1109/access.2023.3269848}
}

@inproceedings{R054,
  author       = {Bettini, Matteo and Shankar, Ajay and Prorok, Amanda},
  title        = {Heterogeneous Multi-Robot Reinforcement Learning},
  booktitle    = {Proc. Int. Conf. Auton. Agents Multiagent Syst. (AAMAS)},
  address      = {London, United Kingdom},
  pages        = {1485--1494},
  year         = {2023},
  month        = {May--Jun.},
  doi          = {10.65109/slob2372}
}

@article{R041,
  author       = {Wheeler, Dylan and Natarajan, Balasubramaniam},
  title        = {Engineering Semantic Communication: A Survey},
  journal      = {IEEE Access},
  volume       = {11},
  pages        = {13965--13995},
  year         = {2023},
  month        = feb,
  doi          = {10.1109/access.2023.3243065}
}

@inproceedings{R068,
  author       = {Kutsevol, Polina and Ayan, Onur and Pappas, Nikolaos and Kellerer, Wolfgang},
  title        = {Goal-Oriented Transport Layer Protocols for Wireless Control},
  booktitle    = {Proc. IEEE Int. Conf. Sens., Commun., Netw. (SECON)},
  address      = {Madrid, Spain},
  pages        = {378--380},
  year         = {2023},
  month        = sep,
  doi          = {10.1109/secon58729.2023.10287494}
}

@article{R042,
  author       = {Stavrou, Photios A. and Kountouris, Marios},
  title        = {The Role of Fidelity in Goal-Oriented Semantic Communication: A Rate Distortion Approach},
  journal      = {IEEE Trans. Commun.},
  volume       = {71},
  number       = {7},
  pages        = {3918--3931},
  year         = {2023},
  month        = jul,
  doi          = {10.1109/tcomm.2023.3274122}
}

@article{R009,
  author       = {Seo, Hyowoon and Park, Jihong and Bennis, Mehdi and Debbah, M{\'e}rouane},
  title        = {Semantics-Native Communication via Contextual Reasoning},
  journal      = {IEEE Trans. Cogn. Commun. Netw.},
  volume       = {9},
  number       = {3},
  pages        = {604--617},
  year         = {2023},
  month        = jun,
  doi          = {10.1109/tccn.2023.3250206}
}

@article{R071,
  author       = {Sheng, Yucheng and Ye, Hao and Liang, Le and Jin, Shi and Li, Geoffrey Ye},
  title        = {Semantic communication for cooperative perception based on importance map},
  journal      = {J. Franklin Inst.},
  volume       = {361},
  number       = {6},
  year         = {2024},
  month        = apr,
  note         = {{Art. no. 106739}},
  doi          = {10.1016/j.jfranklin.2024.106739}
}

@article{R072,
  author       = {Miki, Takahiro and Lee, Joonho and Hwangbo, Jemin and Wellhausen, Lorenz and Koltun, Vladlen and Hutter, Marco},
  title        = {Learning robust perceptive locomotion for quadrupedal robots in the wild},
  journal      = {Sci. Robot.},
  volume       = {7},
  number       = {62},
  year         = {2022},
  month        = jan,
  note         = {{Art. no. eabk2822}},
  doi          = {10.1126/scirobotics.abk2822}
}

@article{R073,
  author       = {Mahato, Prabhat and Saha, Sudipta and Sarkar, Chayan and Shaghil, Md.},
  title        = {Consensus-based fast and energy-efficient multi-robot task allocation},
  journal      = {Robot. Auton. Syst.},
  volume       = {159},
  year         = {2023},
  month        = jan,
  note         = {{Art. no. 104270}},
  doi          = {10.1016/j.robot.2022.104270}
}

@inproceedings{R075,
  author       = {Singh, Sukriti and Srikanthan, Anusha and Mallampati, Vivek and Ravichandar, Harish},
  title        = {Concurrent Constrained Optimization of Unknown Rewards for Multi-Robot Task Allocation},
  booktitle    = {Proc. Robotics: Sci. Syst. (RSS)},
  address      = {Daegu, Republic of Korea},
  year         = {2023},
  month        = jul,
  doi          = {10.15607/rss.2023.xix.108}
}

@article{Jamone2018AffordancesSurvey,
  author       = {Jamone, Lorenzo and Ugur, Emre and Cangelosi, Angelo and Fadiga, Luciano and Bernardino, Alexandre and Piater, Justus and Santos-Victor, Jos{\'e}},
  title        = {Affordances in Psychology, Neuroscience, and Robotics: A Survey},
  journal      = {IEEE Trans. Cogn. Dev. Syst.},
  volume       = {10},
  number       = {1},
  pages        = {4--25},
  year         = {2018},
  month        = mar,
  doi          = {10.1109/tcds.2016.2594134}
}

@article{Mayya2021ResilientTaskAllocation,
  author       = {Mayya, Siddharth and D'Antonio, Diego S. and Salda{\~n}a, David and Kumar, Vijay},
  title        = {Resilient Task Allocation in Heterogeneous Multi-Robot Systems},
  journal      = {IEEE Robot. Autom. Lett.},
  volume       = {6},
  number       = {2},
  pages        = {1327--1334},
  year         = {2021},
  month        = apr,
  doi          = {10.1109/lra.2021.3057559}
}

@article{R013,
  author       = {Yang, Yuwen and Gao, Feifei and Tao, Xiaoming and Liu, Guangyi and Pan, Chengkang},
  title        = {Environment Semantics Aided Wireless Communications: A Case Study of {{mmWave}} Beam Prediction and Blockage Prediction},
  journal      = {IEEE J. Sel. Areas Commun.},
  volume       = {41},
  number       = {7},
  pages        = {2025--2040},
  year         = {2023},
  month        = jul,
  doi          = {10.1109/jsac.2023.3280966}
}

@article{R078,
  author       = {{\c{S}}ahin, Erol and {\c{C}}akmak, Maya and Do{\u{g}}ar, Mehmet R. and U{\u{g}}ur, Emre and {\"U}{\c{c}}oluk, G{\"o}kt{\"u}rk},
  title        = {To Afford or Not to Afford: A New Formalization of Affordances Toward Affordance-Based Robot Control},
  journal      = {Adapt. Behav.},
  volume       = {15},
  number       = {4},
  pages        = {447--472},
  year         = {2007},
  month        = dec,
  doi          = {10.1177/1059712307084689}
}

@article{R084,
  author       = {Salman, Muhammad and Garz{\'o}n Ramos, David and Birattari, Mauro},
  title        = {Automatic design of stigmergy-based behaviours for robot swarms},
  journal      = {Commun. Eng.},
  volume       = {3},
  number       = {1},
  year         = {2024},
  month        = feb,
  note         = {{Art. no. 30}},
  doi          = {10.1038/s44172-024-00175-7}
}

@article{R085,
  author       = {Salman, Muhammad and Garz{\'o}n Ramos, David and Hasselmann, Ken and Birattari, Mauro},
  title        = {{Phormica}: Photochromic Pheromone Release and Detection System for Stigmergic Coordination in Robot Swarms},
  journal      = {Front. Robot. AI},
  volume       = {7},
  year         = {2020},
  month        = dec,
  note         = {{Art. no. 591402}},
  doi          = {10.3389/frobt.2020.591402}
}

@article{R089,
  author       = {Clark, Herbert H. and Wilkes-Gibbs, Deanna},
  title        = {Referring as a collaborative process},
  journal      = {Cognition},
  volume       = {22},
  number       = {1},
  pages        = {1--39},
  year         = {1986},
  month        = feb,
  doi          = {10.1016/0010-0277(86)90010-7}
}

@incollection{R088,
  author       = {Clark, Herbert H. and Brennan, Susan E.},
  title        = {Grounding in communication},
  booktitle    = {Perspectives on Socially Shared Cognition},
  editor       = {Resnick, Lauren B. and Levine, John M. and Teasley, Stephanie D.},
  publisher    = {American Psychological Association},
  address      = {Washington, DC},
  pages        = {127--149},
  year         = {1991},
  doi          = {10.1037/10096-006}
}

@article{R095,
  author       = {Buhrmann, Thomas and Di Paolo, Ezequiel Alejandro and Barandiaran, Xabier},
  title        = {A Dynamical Systems Account of Sensorimotor Contingencies},
  journal      = {Front. Psychol.},
  volume       = {4},
  year         = {2013},
  month        = may,
  note         = {{Art. no. 285}},
  doi          = {10.3389/fpsyg.2013.00285}
}

@article{R096,
  author       = {Yi, Peng and Cao, Yang and Kang, Xin and Liang, Ying-Chang},
  title        = {Deep Learning-Empowered Semantic Communication Systems With a Shared Knowledge Base},
  journal      = {IEEE Trans. Wireless Commun.},
  volume       = {23},
  number       = {6},
  pages        = {6174--6187},
  year         = {2024},
  month        = jun,
  doi          = {10.1109/twc.2023.3330744}
}

@article{R094,
  author       = {Diaz-Calderon, Antonio and Nesnas, Issa A. D. and Nayar, Hari Das and Kim, Won S.},
  title        = {Towards a Unified Representation of Mechanisms for Robotic Control Software},
  journal      = {Int. J. Adv. Robot. Syst.},
  volume       = {3},
  number       = {1},
  pages        = {61--66},
  year         = {2006},
  month        = mar,
  doi          = {10.5772/5757}
}

@article{R097,
  author       = {Olfati-Saber, Reza and Fax, J. Alex and Murray, Richard M.},
  title        = {Consensus and Cooperation in Networked Multi-Agent Systems},
  journal      = {Proc. IEEE},
  volume       = {95},
  number       = {1},
  pages        = {215--233},
  year         = {2007},
  month        = jan,
  doi          = {10.1109/jproc.2006.887293}
}

@article{R098,
  author       = {Gielis, Jennifer and Shankar, Ajay and Prorok, Amanda},
  title        = {A Critical Review of Communications in Multi-robot Systems},
  journal      = {Curr. Robot. Rep.},
  volume       = {3},
  number       = {4},
  pages        = {213--225},
  year         = {2022},
  month        = dec,
  doi          = {10.1007/s43154-022-00090-9}
}

@article{R099,
  author  = {Garc{\'i}a, Paula and Caama{\~n}o, Pilar and Duro, Richard J. and Bellas, Francisco},
  title   = {Scalable Task Assignment for Heterogeneous Multi-Robot Teams},
  journal = {Int. J. Adv. Robot. Syst.},
  volume  = {10},
  number  = {2},
  year    = {2013},
  month   = feb,
  note    = {{Art. no. 105}},
  doi     = {10.5772/55489}
}

@inproceedings{R100,
  author    = {Notomista, Gennaro and Mayya, Siddharth and Hutchinson, Seth and Egerstedt, Magnus},
  title     = {An Optimal Task Allocation Strategy for Heterogeneous Multi-Robot Systems},
  booktitle = {Proc. Eur. Control Conf. (ECC)},
  address   = {Naples, Italy},
  pages     = {2071--2076},
  year      = {2019},
  month     = jun,
  doi       = {10.23919/ecc.2019.8795895}
}

@inproceedings{R101,
  author    = {Lin, Chendi and Luo, Wenhao and Sycara, Katia},
  title     = {Online Connectivity-aware Dynamic Deployment for Heterogeneous Multi-Robot Systems},
  booktitle = {Proc. IEEE Int. Conf. Robot. Autom. (ICRA)},
  address   = {Xi'an, China},
  pages     = {8941--8947},
  year      = {2021},
  month     = {May--Jun.},
  doi       = {10.1109/icra48506.2021.9561748}
}

@inproceedings{R105,
  author    = {Sakaridis, Christos and Dai, Dengxin and Van Gool, Luc},
  title     = {{ACDC}: The Adverse Conditions Dataset with Correspondences for Semantic Driving Scene Understanding},
  booktitle = {Proc. IEEE/CVF Int. Conf. Comput. Vis. (ICCV)},
  address   = {Montreal, QC, Canada},
  pages     = {10745--10755},
  year      = {2021},
  month     = oct,
  doi       = {10.1109/iccv48922.2021.01059}
}

@inproceedings{R106,
  author    = {Hahner, Martin and Sakaridis, Christos and Dai, Dengxin and Van Gool, Luc},
  title     = {Fog Simulation on Real {LiDAR} Point Clouds for {3D} Object Detection in Adverse Weather},
  booktitle = {Proc. IEEE/CVF Int. Conf. Comput. Vis. (ICCV)},
  address   = {Montreal, QC, Canada},
  pages     = {15263--15272},
  year      = {2021},
  month     = oct,
  doi       = {10.1109/iccv48922.2021.01500}
}

@article{R016,
  author  = {Li, Ang and Wei, Xin and Wu, Dan and Zhou, Liang},
  title   = {Cross-Modal Semantic Communications},
  journal = {IEEE Wireless Commun.},
  volume  = {29},
  number  = {6},
  pages   = {144--151},
  year    = {2022},
  month   = dec,
  doi     = {10.1109/mwc.008.2200180}
}

@inproceedings{R117,
  author    = {Bai, Xuyang and Hu, Zeyu and Zhu, Xinge and Huang, Qingqiu and Chen, Yilun and Fu, Hongbo and Tai, Chiew-Lan},
  title     = {{TransFusion}: Robust {LiDAR}-Camera Fusion for {3D} Object Detection with Transformers},
  booktitle = {Proc. IEEE/CVF Conf. Comput. Vis. Pattern Recognit. (CVPR)},
  address   = {New Orleans, LA, USA},
  pages     = {1080--1089},
  year      = {2022},
  month     = jun,
  doi       = {10.1109/cvpr52688.2022.00116}
}

@inproceedings{R120,
  author    = {Zhang, Yanan and Chen, Jiaxin and Huang, Di},
  title     = {{CAT-Det}: Contrastively Augmented Transformer for Multimodal {3D} Object Detection},
  booktitle = {Proc. IEEE/CVF Conf. Comput. Vis. Pattern Recognit. (CVPR)},
  address   = {New Orleans, LA, USA},
  pages     = {898--907},
  year      = {2022},
  month     = jun,
  doi       = {10.1109/cvpr52688.2022.00098}
}

@article{R121,
  author  = {Zhang, Jiaming and Liu, Huayao and Yang, Kailun and Hu, Xinxin and Liu, Ruiping and Stiefelhagen, Rainer},
  title   = {{CMX}: Cross-Modal Fusion for {RGB-X} Semantic Segmentation With Transformers},
  journal = {IEEE Trans. Intell. Transp. Syst.},
  volume  = {24},
  number  = {12},
  pages   = {14679--14694},
  year    = {2023},
  month   = dec,
  doi     = {10.1109/tits.2023.3300537}
}

@inproceedings{R129,
  author    = {Girdhar, Rohit and El-Nouby, Alaaeldin and Liu, Zhuang and Singh, Mannat and Alwala, Kalyan Vasudev and Joulin, Armand and Misra, Ishan},
  title     = {{ImageBind} One Embedding Space to Bind Them All},
  booktitle = {Proc. IEEE/CVF Conf. Comput. Vis. Pattern Recognit. (CVPR)},
  address   = {Vancouver, BC, Canada},
  pages     = {15180--15190},
  year      = {2023},
  month     = jun,
  doi       = {10.1109/cvpr52729.2023.01457}
}

@inproceedings{R113,
  author    = {Palladin, Edoardo and Dietze, Roland and Narayanan, Praveen and Bijelic, Mario and Heide, Felix},
  title     = {{SAMFusion}: Sensor-Adaptive Multimodal Fusion for {3D} Object Detection in Adverse Weather},
  booktitle = {Proc. Eur. Conf. Comput. Vis. (ECCV)},
  address   = {Milan, Italy},
  pages     = {484--503},
  year      = {2024},
  month     = {Sep.--Oct.},
  doi       = {10.1007/978-3-031-73030-6_27}
}

@inproceedings{R123,
  author    = {Ge, Chongjian and Chen, Junsong and Xie, Enze and Wang, Zhongdao and Hong, Lanqing and Lu, Huchuan and Li, Zhenguo and Luo, Ping},
  title     = {{MetaBEV}: Solving Sensor Failures for {3D} Detection and Map Segmentation},
  booktitle = {Proc. IEEE/CVF Int. Conf. Comput. Vis. (ICCV)},
  address   = {Paris, France},
  pages     = {8687--8697},
  year      = {2023},
  month     = oct,
  doi       = {10.1109/iccv51070.2023.00801}
}

@article{R124,
  author  = {Chen, Siran and Ma, Yue and Qiao, Yu and Wang, Yali},
  title   = {{M-BEV}: Masked {BEV} Perception for Robust Autonomous Driving},
  journal = {Proc. AAAI Conf. Artif. Intell.},
  volume  = {38},
  number  = {2},
  pages   = {1183--1191},
  year    = {2024},
  month   = mar,
  doi     = {10.1609/aaai.v38i2.27880}
}

@article{R126,
  author  = {Huang, Xun and Xu, Ziyu and Wu, Hai and Wang, Jinlong and Xia, Qiming and Xia, Yan and Li, Jonathan and Gao, Kyle and Wen, Chenglu and Wang, Cheng},
  title   = {{L4DR}: {LiDAR}-{4DRadar} Fusion for Weather-Robust {3D} Object Detection},
  journal = {Proc. AAAI Conf. Artif. Intell.},
  volume  = {39},
  number  = {4},
  pages   = {3806--3814},
  year    = {2025},
  month   = apr,
  doi     = {10.1609/aaai.v39i4.32397}
}

@inproceedings{R127,
  author    = {Jaritz, Maximilian and Vu, Tuan-Hung and de Charette, Raoul and Wirbel, Emilie and Perez, Patrick},
  title     = {{xMUDA}: Cross-Modal Unsupervised Domain Adaptation for {3D} Semantic Segmentation},
  booktitle = {Proc. IEEE/CVF Conf. Comput. Vis. Pattern Recognit. (CVPR)},
  address   = {Seattle, WA, USA},
  pages     = {12602--12611},
  year      = {2020},
  month     = jun,
  doi       = {10.1109/cvpr42600.2020.01262}
}

@inproceedings{R115,
  author    = {Lee, Yi-Lun and Tsai, Yi-Hsuan and Chiu, Wei-Chen and Lee, Chen-Yu},
  title     = {Multimodal Prompting with Missing Modalities for Visual Recognition},
  booktitle = {Proc. IEEE/CVF Conf. Comput. Vis. Pattern Recognit. (CVPR)},
  address   = {Vancouver, BC, Canada},
  pages     = {14943--14952},
  year      = {2023},
  month     = jun,
  doi       = {10.1109/cvpr52729.2023.01435}
}

@inproceedings{R128,
  author    = {Wang, Hu and Chen, Yuanhong and Ma, Congbo and Avery, Jodie and Hull, Louise and Carneiro, Gustavo},
  title     = {Multi-Modal Learning with Missing Modality via Shared-Specific Feature Modelling},
  booktitle = {Proc. IEEE/CVF Conf. Comput. Vis. Pattern Recognit. (CVPR)},
  address   = {Vancouver, BC, Canada},
  pages     = {15878--15887},
  year      = {2023},
  month     = jun,
  doi       = {10.1109/cvpr52729.2023.01524}
}

@article{R116,
  author  = {Liu, Yinqiu and Du, Hongyang and Niyato, Dusit and Kang, Jiawen and Xiong, Zehui and Mao, Shiwen and Zhang, Ping and Shen, Xuemin},
  title   = {Cross-Modal Generative Semantic Communications for Mobile {AIGC}: Joint Semantic Encoding and Prompt Engineering},
  journal = {IEEE Trans. Mobile Comput.},
  volume  = {23},
  number  = {12},
  pages   = {14871--14888},
  year    = {2024},
  month   = dec,
  doi     = {10.1109/tmc.2024.3449645}
}

@inproceedings{R132,
  author    = {Xu, Runsheng and Xiang, Hao and Tu, Zhengzhong and Xia, Xin and Yang, Ming-Hsuan and Ma, Jiaqi},
  title     = {{V2X-ViT}: Vehicle-to-Everything Cooperative Perception with Vision Transformer},
  booktitle = {Proc. Eur. Conf. Comput. Vis. (ECCV)},
  address   = {Tel Aviv, Israel},
  pages     = {107--124},
  year      = {2022},
  month     = oct,
  doi       = {10.1007/978-3-031-19842-7_7}
}

@inproceedings{R136,
  author    = {Xu, Runsheng and Tu, Zhengzhong and Xiang, Hao and Shao, Wei and Zhou, Bolei and Ma, Jiaqi},
  title     = {{CoBEVT}: Cooperative Bird's Eye View Semantic Segmentation with Sparse Transformers},
  booktitle = {Proc. Conf. Robot Learn. (CoRL)},
  address   = {Auckland, New Zealand},
  pages     = {989--1000},
  year      = {2022},
  month     = dec
}

@inproceedings{R137,
  author    = {Yuan, Yunshuang and Xia, Yan and Cremers, Daniel and Sester, Monika},
  title     = {{SparseAlign}: A Fully Sparse Framework for Cooperative Object Detection},
  booktitle = {Proc. IEEE/CVF Conf. Comput. Vis. Pattern Recognit. (CVPR)},
  address   = {Nashville, TN, USA},
  pages     = {22296--22305},
  year      = {2025},
  month     = jun,
  doi       = {10.1109/cvpr52734.2025.02077}
}

@inproceedings{R133,
  author    = {Wang, Tianhang and Chen, Guang and Chen, Kai and Liu, Zhengfa and Zhang, Bo and Knoll, Alois and Jiang, Changjun},
  title     = {{UMC}: A Unified Bandwidth-efficient and Multi-resolution based Collaborative Perception Framework},
  booktitle = {Proc. IEEE/CVF Int. Conf. Comput. Vis. (ICCV)},
  address   = {Paris, France},
  pages     = {8153--8162},
  year      = {2023},
  month     = oct,
  doi       = {10.1109/iccv51070.2023.00752}
}

@inproceedings{R138,
  author    = {Liu, Yen-Cheng and Tian, Junjiao and Glaser, Nathaniel and Kira, Zsolt},
  title     = {{When2com}: Multi-Agent Perception via Communication Graph Grouping},
  booktitle = {Proc. IEEE/CVF Conf. Comput. Vis. Pattern Recognit. (CVPR)},
  address   = {Seattle, WA, USA},
  pages     = {4105--4114},
  year      = {2020},
  month     = jun,
  doi       = {10.1109/cvpr42600.2020.00416}
}

@inproceedings{R139,
  author    = {Liu, Yen-Cheng and Tian, Junjiao and Ma, Chih-Yao and Glaser, Nathan and Kuo, Chia-Wen and Kira, Zsolt},
  title     = {{Who2com}: Collaborative Perception via Learnable Handshake Communication},
  booktitle = {Proc. IEEE Int. Conf. Robot. Autom. (ICRA)},
  address   = {Paris, France},
  pages     = {6876--6883},
  year      = {2020},
  month     = {May--Aug.},
  doi       = {10.1109/icra40945.2020.9197364}
}

@article{R140,
  author  = {Luo, Guiyang and Shao, Chongzhang and Cheng, Nan and Zhou, Haibo and Zhang, Hui and Yuan, Quan and Li, Jinglin},
  title   = {{EdgeCooper}: Network-Aware Cooperative {LiDAR} Perception for Enhanced Vehicular Awareness},
  journal = {IEEE J. Sel. Areas Commun.},
  volume  = {42},
  number  = {1},
  pages   = {207--222},
  year    = {2024},
  month   = jan,
  doi     = {10.1109/jsac.2023.3322764}
}

@inproceedings{R134,
  author    = {Yu, Haibao and Luo, Yizhen and Shu, Mao and Huo, Yiyi and Yang, Zebang and Shi, Yifeng and Guo, Zhenglong and Li, Hanyu and Hu, Xing and Yuan, Jirui and Nie, Zaiqing},
  title     = {{DAIR-V2X}: A Large-Scale Dataset for Vehicle-Infrastructure Cooperative {3D} Object Detection},
  booktitle = {Proc. IEEE/CVF Conf. Comput. Vis. Pattern Recognit. (CVPR)},
  address   = {New Orleans, LA, USA},
  pages     = {21329--21338},
  year      = {2022},
  month     = jun,
  doi       = {10.1109/cvpr52688.2022.02067}
}

@article{R135,
  author  = {Yu, Haibao and Yang, Wenxian and Zhong, Jiaru and Yang, Zhenwei and Fan, Siqi and Luo, Ping and Nie, Zaiqing},
  title   = {End-to-End Autonomous Driving Through {V2X} Cooperation},
  journal = {Proc. AAAI Conf. Artif. Intell.},
  volume  = {39},
  number  = {9},
  pages   = {9598--9606},
  year    = {2025},
  month   = apr,
  doi     = {10.1609/aaai.v39i9.33040}
}

@inproceedings{R141,
  author    = {Xu, Runsheng and Xiang, Hao and Xia, Xin and Han, Xu and Li, Jinlong and Ma, Jiaqi},
  title     = {{OPV2V}: An Open Benchmark Dataset and Fusion Pipeline for Perception with Vehicle-to-Vehicle Communication},
  booktitle = {Proc. IEEE Int. Conf. Robot. Autom. (ICRA)},
  address   = {Philadelphia, PA, USA},
  pages     = {2583--2589},
  year      = {2022},
  month     = may,
  doi       = {10.1109/icra46639.2022.9812038}
}

@inproceedings{R142,
  author    = {Yu, Haibao and Yang, Wenxian and Ruan, Hongzhi and Yang, Zhenwei and Tang, Yingjuan and Gao, Xu and Hao, Xin and Shi, Yifeng and Pan, Yifeng and Sun, Ning and Song, Juan and Yuan, Jirui and Luo, Ping and Nie, Zaiqing},
  title     = {{V2X-Seq}: A Large-Scale Sequential Dataset for Vehicle-Infrastructure Cooperative Perception and Forecasting},
  booktitle = {Proc. IEEE/CVF Conf. Comput. Vis. Pattern Recognit. (CVPR)},
  address   = {Vancouver, BC, Canada},
  pages     = {5486--5495},
  year      = {2023},
  month     = jun,
  doi       = {10.1109/cvpr52729.2023.00531}
}

@inproceedings{R144,
  author    = {Lu, Yifan and Li, Quanhao and Liu, Baoan and Dianati, Mehrdad and Feng, Chen and Chen, Siheng and Wang, Yanfeng},
  title     = {Robust Collaborative {3D} Object Detection in Presence of Pose Errors},
  booktitle = {Proc. IEEE Int. Conf. Robot. Autom. (ICRA)},
  address   = {London, United Kingdom},
  pages     = {4812--4818},
  year      = {2023},
  month     = {May--Jun.},
  doi       = {10.1109/icra48891.2023.10160546}
}

@inproceedings{R143,
  author    = {Su, Sanbao and Li, Yiming and He, Sihong and Han, Songyang and Feng, Chen and Ding, Caiwen and Miao, Fei},
  title     = {Uncertainty Quantification of Collaborative Detection for Self-Driving},
  booktitle = {Proc. IEEE Int. Conf. Robot. Autom. (ICRA)},
  address   = {London, United Kingdom},
  pages     = {5588--5594},
  year      = {2023},
  month     = {May--Jun.},
  doi       = {10.1109/icra48891.2023.10160367}
}

@article{R145,
  author  = {Ren, Shunli and Lei, Zixing and Wang, Zi and Dianati, Mehrdad and Wang, Yafei and Chen, Siheng and Zhang, Wenjun},
  title   = {Interruption-Aware Cooperative Perception for {V2X} Communication-Aided Autonomous Driving},
  journal = {IEEE Trans. Intell. Veh.},
  volume  = {9},
  number  = {4},
  pages   = {4698--4714},
  year    = {2024},
  month   = apr,
  doi     = {10.1109/tiv.2024.3371974}
}

@inproceedings{R103,
  author    = {Xia, Le and Sun, Yao and Li, Xiaoqian and Feng, Gang and Imran, Muhammad Ali},
  title     = {Wireless Resource Management in Intelligent Semantic Communication Networks},
  booktitle = {Proc. IEEE INFOCOM Workshops (INFOCOM WKSHPS)},
  address   = {Virtual Conf.},
  pages     = {1--6},
  year      = {2022},
  month     = may,
  doi       = {10.1109/infocomwkshps54753.2022.9797984}
}

@article{R146,
  author  = {Mena, Juan P. and N{\'u}{\~n}ez, Felipe},
  title   = {Age of Information in {IoT}-Based Networked Control Systems: A {MAC} Perspective},
  journal = {Automatica},
  volume  = {147},
  year    = {2023},
  month   = jan,
  note    = {{Art. no. 110652}},
  doi     = {10.1016/j.automatica.2022.110652}
}

@article{R148,
  author  = {Durisi, Giuseppe and Koch, Tobias and Popovski, Petar},
  title   = {Toward Massive, Ultrareliable, and Low-Latency Wireless Communication With Short Packets},
  journal = {Proc. IEEE},
  volume  = {104},
  number  = {9},
  pages   = {1711--1726},
  year    = {2016},
  month   = sep,
  doi     = {10.1109/jproc.2016.2537298}
}

@article{R020,
  author  = {Fan, Senran and Liang, Haotai and Dong, Chen and Xu, Xiaodong and Liu, Geng},
  title   = {A Specific Task-Oriented Semantic Image Communication System for Substation Patrol Inspection},
  journal = {IEEE Trans. Power Del.},
  volume  = {39},
  number  = {2},
  pages   = {835--844},
  year    = {2024},
  month   = apr,
  doi     = {10.1109/tpwrd.2023.3337274}
}

@article{R151,
  author  = {He, Yangshuo and Yu, Guanding and Cai, Yunlong},
  title   = {Rate-Adaptive Coding Mechanism for Semantic Communications With Multi-Modal Data},
  journal = {IEEE Trans. Commun.},
  volume  = {72},
  number  = {3},
  pages   = {1385--1400},
  year    = {2024},
  month   = mar,
  doi     = {10.1109/tcomm.2023.3335977}
}

@inproceedings{R155,
  author    = {Ma, Yi and Xu, Chunmei and Liu, Zhenyu and Zhang, Siqi and Tafazolli, Rahim},
  title     = {Importance-Aware Source-Channel Coding for Multi-Modal Task-Oriented Semantic Communication},
  booktitle = {Proc. IEEE Int. Conf. Mach. Learn. Commun. Netw. (ICMLCN)},
  address   = {Barcelona, Spain},
  pages     = {1--6},
  year      = {2025},
  month     = may,
  doi       = {10.1109/icmlcn64995.2025.11140558}
}

@article{R156,
  author  = {Sun, Zhiye and Ma, Shuai and Li, Shiyin},
  title   = {Task-Oriented Semantic Communication With Importance-Aware Rate Control},
  journal = {IEEE Commun. Lett.},
  volume  = {29},
  number  = {7},
  pages   = {1520--1524},
  year    = {2025},
  month   = jul,
  doi     = {10.1109/lcomm.2025.3566092}
}

@article{R152,
  author  = {Zhang, Guangyi and Li, Hanlei and Cai, Yunlong and Hu, Qiyu and Yu, Guanding and Qin, Zhijin},
  title   = {Progressive Learned Image Transmission for Semantic Communication Using Hierarchical {VAE}},
  journal = {IEEE Trans. Cogn. Commun. Netw.},
  volume  = {11},
  number  = {6},
  pages   = {3640--3654},
  year    = {2025},
  month   = dec,
  doi     = {10.1109/tccn.2025.3546935}
}

@inproceedings{R157,
  author    = {Nam, Hyelin and Park, Jihong and Choi, Jinho and Kim, Seong-Lyun},
  title     = {Sequential Semantic Generative Communication for Progressive Text-to-Image Generation},
  booktitle = {Proc. IEEE Int. Conf. Sens., Commun., Netw. (SECON)},
  address   = {Madrid, Spain},
  pages     = {91--94},
  year      = {2023},
  month     = sep,
  doi       = {10.1109/secon58729.2023.10287475}
}

@inproceedings{R158,
  author    = {Riherd, Eva and Mudumbai, Raghu and Xu, Weiyu},
  title     = {Linear Progressive Coding for Semantic Communication using Deep Neural Networks},
  booktitle = {Proc. Annu. Conf. Inf. Sci. Syst. (CISS)},
  address   = {Princeton, NJ, USA},
  pages     = {1--6},
  year      = {2024},
  month     = mar,
  doi       = {10.1109/ciss59072.2024.10480188}
}

@article{R153,
  author  = {Jiang, Peiwen and Wen, Chao-Kai and Jin, Shi and Li, Geoffrey Ye},
  title   = {Deep Source-Channel Coding for Sentence Semantic Transmission With {HARQ}},
  journal = {IEEE Trans. Commun.},
  volume  = {70},
  number  = {8},
  pages   = {5225--5240},
  year    = {2022},
  month   = aug,
  doi     = {10.1109/tcomm.2022.3180997}
}

@article{R159,
  author  = {Zhou, Qingyang and Li, Rongpeng and Zhao, Zhifeng and Xiao, Yong and Zhang, Honggang},
  title   = {Adaptive Bit Rate Control in Semantic Communication With Incremental Knowledge-Based {HARQ}},
  journal = {IEEE Open J. Commun. Soc.},
  volume  = {3},
  pages   = {1076--1089},
  year    = {2022},
  month   = jul,
  doi     = {10.1109/ojcoms.2022.3189023}
}

@article{R160,
  author  = {Ni, Fei and Li, Rongpeng and Zhao, Zhifeng and Zhang, Honggang},
  title   = {Topology data analysis-based error detection for semantic image transmission with incremental knowledge-based {HARQ}},
  journal = {China Commun.},
  volume  = {22},
  number  = {1},
  pages   = {235--255},
  year    = {2025},
  month   = jan,
  doi     = {10.23919/jcc.fa.2023-0799.202501}
}

@inproceedings{R165,
  author    = {Kendall, Alex and Gal, Yarin},
  title     = {What Uncertainties Do We Need in {Bayesian} Deep Learning for Computer Vision?},
  booktitle = {Proc. Adv. Neural Inf. Process. Syst. (NeurIPS)},
  address   = {Long Beach, CA, USA},
  pages     = {5574--5584},
  year      = {2017},
  month     = dec
}

@article{Ren2024KnowledgeBaseSemanticCommunication,
  author  = {Ren, Jinke and Zhang, Zezhong and Xu, Jie and Chen, Guanying and Sun, Yaping and Zhang, Ping and Cui, Shuguang},
  title   = {Knowledge Base Enabled Semantic Communication: A Generative Perspective},
  journal = {IEEE Wireless Commun.},
  volume  = {31},
  number  = {4},
  pages   = {14--22},
  year    = {2024},
  month   = aug,
  doi     = {10.1109/mwc.001.2300553}
}

@inproceedings{Ghasemi2020TaskOrientedActivePerception,
  author    = {Ghasemi, Mahsa and Bulgur, Erdem and Topcu, Ufuk},
  title     = {Task-Oriented Active Perception and Planning in Environments with Partially Known Semantics},
  booktitle = {Proc. Int. Conf. Mach. Learn. (ICML)},
  address   = {Virtual Conf.},
  pages     = {3484--3493},
  year      = {2020},
  month     = jul
}

@article{Paulius2019KnowledgeRepresentationRobotics,
  author  = {Paulius, David and Sun, Yu},
  title   = {A Survey of Knowledge Representation in Service Robotics},
  journal = {Robot. Auton. Syst.},
  volume  = {118},
  pages   = {13--30},
  year    = {2019},
  month   = aug,
  doi     = {10.1016/j.robot.2019.03.005}
}

@inproceedings{R183,
  author    = {Chisari, Eugenio and von Hartz, Jan Ole and Despinoy, Fabien and Valada, Abhinav},
  title     = {Robotic Task Ambiguity Resolution via Natural Language Interaction},
  booktitle = {Proc. IEEE/RSJ Int. Conf. Intell. Robots Syst. (IROS)},
  address   = {Hangzhou, China},
  pages     = {14821--14827},
  year      = {2025},
  month     = oct,
  doi       = {10.1109/iros60139.2025.11247661}
}

@inproceedings{R184,
  author    = {Kuehn, Hannah and Santos, Leonardo and Leite, Iolanda},
  title     = {Clarifying Constraints in Interactive Robot Learning with Language Feedback},
  booktitle = {Proc. ACM/IEEE Int. Conf. Human-Robot Interact. (HRI)},
  address   = {Edinburgh, United Kingdom},
  pages     = {816--824},
  year      = {2026},
  month     = mar,
  doi       = {10.1145/3757279.3785583}
}

@article{NiuZhang2024SemanticTheory,
  author  = {Niu, Kai and Zhang, Ping},
  title   = {A Mathematical Theory of Semantic Communication},
  journal = {J. Commun.},
  volume  = {45},
  number  = {6},
  pages   = {7--59},
  year    = {2024},
  month   = jun,
  doi     = {10.11959/j.issn.1000-436x.2024111}
}

@book{R193,
  author    = {Niu, Kai and Zhang, Ping},
  title     = {The Mathematical Theory of Semantic Communication},
  address   = {Singapore},
  publisher = {Springer},
  year      = {2025},
  doi       = {10.1007/978-981-96-5132-0},
  url       = {https://link.springer.com/book/10.1007/978-981-96-5132-0}
}

@inproceedings{Ha2018WorldModels,
  author    = {Ha, David and Schmidhuber, J{\"u}rgen},
  title     = {Recurrent {World Models} Facilitate Policy Evolution},
  booktitle = {Proc. Adv. Neural Inf. Process. Syst. (NeurIPS)},
  address   = {Montr{\'e}al, QC, Canada},
  pages     = {2450--2462},
  year      = {2018},
  month     = dec
}

@misc{LeCun2022Path,
  author       = {LeCun, Yann},
  title        = {A Path Towards Autonomous Machine Intelligence},
  howpublished = {OpenReview position paper},
  note         = {Version 0.9.2},
  year         = {2022},
  month        = jun,
  url          = {https://openreview.net/forum?id=BZ5a1r-kVsf}
}

@inproceedings{Assran2023IJEPA,
  author    = {Assran, Mahmoud and Duval, Quentin and Misra, Ishan and Bojanowski, Piotr and Vincent, Pascal and Rabbat, Michael and LeCun, Yann and Ballas, Nicolas},
  title     = {Self-Supervised Learning from Images with a Joint-Embedding Predictive Architecture},
  booktitle = {Proc. IEEE/CVF Conf. Comput. Vis. Pattern Recognit. (CVPR)},
  address   = {Vancouver, BC, Canada},
  pages     = {15619--15629},
  year      = {2023},
  month     = jun,
  doi       = {10.1109/cvpr52729.2023.01499}
}

@inproceedings{Wu2024iVideoGPT,
  author    = {Wu, Jialong and Yin, Shaofeng and Feng, Ningya and He, Xu and Li, Dong and Hao, Jianye and Long, Mingsheng},
  title     = {{iVideoGPT}: Interactive {VideoGPTs} are Scalable {World Models}},
  booktitle = {Proc. Adv. Neural Inf. Process. Syst. (NeurIPS)},
  address   = {Vancouver, BC, Canada},
  pages     = {68082--68119},
  year      = {2024},
  month     = dec,
  doi       = {10.52202/079017-2173}
}

@inproceedings{Zhu2025IRASim,
  author    = {Zhu, Fangqi and Wu, Hongtao and Guo, Song and Liu, Yuxiao and Cheang, Chilam and Kong, Tao},
  title     = {{IRASim}: A Fine-Grained {World Model} for Robot Manipulation},
  booktitle = {Proc. IEEE/CVF Int. Conf. Comput. Vis. (ICCV)},
  address   = {Honolulu, HI, USA},
  pages     = {9834--9844},
  year      = {2025},
  month     = oct,
  doi       = {10.1109/iccv51701.2025.00917}
}

@article{Yang2024STP,
  author  = {Yang, Jiange and Liu, Bei and Fu, Jianlong and Pan, Bocheng and Wu, Gangshan and Wang, Limin},
  title   = {Spatiotemporal Predictive Pre-training for Robotic Motor Control},
  journal = {Int. J. Comput. Vis.},
  volume  = {134},
  number  = {8},
  year    = {2026},
  month   = aug,
  note    = {{Art. no. 354}},
  doi     = {10.1007/s11263-026-02948-3}
}

@misc{Burchi2024MuDreamer,
  author        = {Burchi, Maxime and Timofte, Radu},
  title         = {{MuDreamer}: Learning Predictive {World Models} without Reconstruction},
  year          = {2024},
  month         = may,
  howpublished  = {arXiv:2405.15083},
  eprint        = {2405.15083},
  archivePrefix = {arXiv},
  primaryClass  = {cs.AI},
  doi           = {10.48550/arxiv.2405.15083},
  url           = {https://arxiv.org/abs/2405.15083}
}

@inproceedings{Lim2022Real2Sim2Real,
  author    = {Lim, Vincent and Huang, Huang and Chen, Lawrence Yunliang and Wang, Jonathan and Ichnowski, Jeffrey and Seita, Daniel and Laskey, Michael and Goldberg, Ken},
  title     = {{Real2Sim2Real}: Self-Supervised Learning of Physical Single-Step Dynamic Actions for Planar Robot Casting},
  booktitle = {Proc. IEEE Int. Conf. Robot. Autom. (ICRA)},
  address   = {Philadelphia, PA, USA},
  pages     = {8282--8289},
  year      = {2022},
  month     = may,
  doi       = {10.1109/icra46639.2022.9811651}
}

@misc{Wong2023WordToWorld,
  author        = {Wong, Lionel and Grand, Gabriel and Lew, Alexander K. and Goodman, Noah D. and Mansinghka, Vikash K. and Andreas, Jacob and Tenenbaum, Joshua B.},
  title         = {From Word Models to {World Models}: Translating from Natural Language to the Probabilistic Language of Thought},
  year          = {2023},
  month         = jun,
  howpublished  = {arXiv:2306.12672},
  eprint        = {2306.12672},
  archivePrefix = {arXiv},
  primaryClass  = {cs.CL},
  doi           = {10.48550/arxiv.2306.12672},
  url           = {https://arxiv.org/abs/2306.12672}
}

@article{R232,
  author  = {Ross, St{\'e}phane and Pineau, Joelle and Paquet, S{\'e}bastien and Chaib-draa, Brahim},
  title   = {Online Planning Algorithms for {POMDPs}},
  journal = {J. Artif. Intell. Res.},
  volume  = {32},
  pages   = {663--704},
  year    = {2008},
  month   = jul,
  doi     = {10.1613/jair.2567}
}

@article{R231,
  author  = {Smallwood, Richard D. and Sondik, Edward J.},
  title   = {The Optimal Control of Partially Observable {Markov} Processes over a Finite Horizon},
  journal = {Oper. Res.},
  volume  = {21},
  number  = {5},
  pages   = {1071--1088},
  year    = {1973},
  month   = {Sep./Oct.},
  doi     = {10.1287/opre.21.5.1071}
}

@inproceedings{Lowe2017MADDPG,
  author    = {Lowe, Ryan and Wu, Yi and Tamar, Aviv and Harb, Jean and Abbeel, Pieter and Mordatch, Igor},
  title     = {Multi-Agent Actor-Critic for Mixed Cooperative-Competitive Environments},
  booktitle = {Proc. Adv. Neural Inf. Process. Syst. (NeurIPS)},
  address   = {Long Beach, CA, USA},
  pages     = {6379--6390},
  year      = {2017},
  month     = dec
}

@inproceedings{Rashid2018QMIX,
  author    = {Rashid, Tabish and Samvelyan, Mikayel and Schroeder de Witt, Christian and Farquhar, Gregory and Foerster, Jakob and Whiteson, Shimon},
  title     = {{QMIX}: Monotonic Value Function Factorisation for Deep Multi-Agent Reinforcement Learning},
  booktitle = {Proc. Int. Conf. Mach. Learn. (ICML)},
  address   = {Stockholm, Sweden},
  pages     = {4295--4304},
  year      = {2018},
  month     = jul
}

@inproceedings{R244,
  author    = {Das, Abhishek and Gervet, Th{\'e}ophile and Romoff, Joshua and Batra, Dhruv and Parikh, Devi and Rabbat, Mike and Pineau, Joelle},
  title     = {{TarMAC}: Targeted Multi-Agent Communication},
  booktitle = {Proc. Int. Conf. Mach. Learn. (ICML)},
  address   = {Long Beach, CA, USA},
  pages     = {1538--1546},
  year      = {2019},
  month     = jun
}

@article{Liang2026WirelessEmbodiedIntelligence,
  author  = {Liang, Hongtao and Diao, Yihe and Wu, Yuhang and Zhou, Fuhui and Wu, Qihui},
  title   = {Synergetic Empowerment: Wireless Communications Meets Embodied Intelligence},
  journal = {IEEE Commun. Mag.},
  year    = {2026},
  month   = jun,
  note    = {early access},
  doi     = {10.1109/mcom.001.2500569}
}

@article{Chaccour2025LessDataMoreKnowledge,
  author  = {Chaccour, Christina and Saad, Walid and Debbah, M{\'e}rouane and Han, Zhu and Poor, H. Vincent},
  title   = {Less Data, More Knowledge: Building Next-Generation Semantic Communication Networks},
  journal = {IEEE Commun. Surveys Tuts.},
  volume  = {27},
  number  = {1},
  pages   = {37--76},
  year    = {2025},
  month   = feb,
  doi     = {10.1109/comst.2024.3412852}
}

@article{Lu2024SemanticsEmpoweredSurvey,
  author  = {Lu, Zhilin and Li, Rongpeng and Lu, Kun and Chen, Xianfu and Hossain, Ekram and Zhao, Zhifeng and Zhang, Honggang},
  title   = {Semantics-Empowered Communications: A Tutorial-Cum-Survey},
  journal = {IEEE Commun. Surveys Tuts.},
  volume  = {26},
  number  = {1},
  pages   = {41--79},
  year    = {2024},
  month   = {1st Quart.},
  doi     = {10.1109/comst.2023.3333342}
}

@article{Zhang2025IntelliciseWirelessSurvey,
  author  = {Zhang, Ping and Xu, Wenjun and Liu, Yiming and Qin, Xiaoqi and Niu, Kai and Cui, Shuguang and Shi, Guangming and Qin, Zhijin and Xu, Xiaodong and Wang, Fengyu and Meng, Yue and Dong, Chen and Dai, Jincheng and Yang, Qianqian and Sun, Yaping and Gao, Dahua and Gao, Hui and Han, Shujun and Song, Xiaodan},
  title   = {Intellicise Wireless Networks From Semantic Communications: A Survey, Research Issues, and Challenges},
  journal = {IEEE Commun. Surveys Tuts.},
  volume  = {27},
  number  = {3},
  pages   = {2051--2084},
  year    = {2025},
  month   = jun,
  doi     = {10.1109/comst.2024.3443193}
}

@article{Liang2025GenerativeAISemanticSurvey,
  author  = {Liang, Chengsi and Du, Hongyang and Sun, Yao and Niyato, Dusit and Kang, Jiawen and Zhao, Dezong and Imran, Muhammad Ali},
  title   = {Generative {AI}-Driven Semantic Communication Networks: Architecture, Technologies, and Applications},
  journal = {IEEE Trans. Cogn. Commun. Netw.},
  volume  = {11},
  number  = {1},
  pages   = {27--47},
  year    = {2025},
  month   = feb,
  doi     = {10.1109/tccn.2024.3435524}
}

@article{Zhang2026ComAI,
  author  = {Zhang, Ping and Niu, Kai and Wang, Xiaoyun and Liu, Yiming and Liang, Zijian and Dong, Chen and Dai, Jincheng and Xu, Xiaodong and Xu, Wenjun and Zhang, Zhi and Wang, Guangyu and Li, Yanlu and Wu, Di and Wu, Hequan},
  title   = {{ComAI}: The Convergence of Communication and Artificial Intelligence},
  journal = {IEEE Commun. Surveys Tuts.},
  volume  = {28},
  pages   = {2163--2197},
  year    = {2026},
  doi     = {10.1109/comst.2025.3608174}
}

@article{Zhu2026CognitiveDigitalTwinSurvey,
  author  = {Zhu, Fang and Chen, Jiayuan and Wen, Junjie and Yang, Yuye and Yi, Changyan and Tie, Yun and Zhang, Peng and Cai, Jun and Niyato, Dusit and Guizani, Mohsen},
  title   = {From Data Mirror to Smart Copilot: A Survey on {NextG} Semantic Communication for Propelling Digital Twin World Into Cognitive Stage},
  journal = {IEEE Commun. Surveys Tuts.},
  volume  = {28},
  pages   = {4915--4947},
  year    = {2026},
  doi     = {10.1109/comst.2026.3665395}
}

@article{Feng2025SemanticAutonomousDriving,
  author  = {Feng, Yunqi and Shen, Hesheng and Shan, Zhendong and Yang, Qianqian and Shi, Xiufang},
  title   = {Semantic Communication for Edge Intelligence Enabled Autonomous Driving System},
  journal = {IEEE Netw.},
  volume  = {39},
  number  = {2},
  pages   = {149--157},
  year    = {2025},
  month   = mar,
  doi     = {10.1109/mnet.2024.3468328}
}

@article{Sagduyu2024JointSensingSemantic,
  author  = {Sagduyu, Yalin E. and Erpek, Tugba and Yener, Aylin and Ulukus, Sennur},
  title   = {Joint Sensing and Semantic Communications with Multi-Task Deep Learning},
  journal = {IEEE Commun. Mag.},
  volume  = {62},
  number  = {9},
  pages   = {74--81},
  year    = {2024},
  month   = sep,
  doi     = {10.1109/mcom.002.2300640}
}

@article{Li2025EmbodiedMultiAgentSystems,
  author  = {Li, Zhuo and Wu, Weiran and Guo, Yunlong and Sun, Jian and Han, Qing-Long},
  title   = {Embodied Multi-Agent Systems: A Review},
  journal = {IEEE/CAA J. Autom. Sinica},
  volume  = {12},
  number  = {6},
  pages   = {1095--1116},
  year    = {2025},
  month   = jun,
  doi     = {10.1109/jas.2025.125552}
}

@article{KA2024MRTAReview,
  author    = {A., Athira K. and Udayan, J. Divya and Subramaniam, Umashankar},
  title     = {A Systematic Literature Review on Multi-Robot Task Allocation},
  journal   = {ACM Comput. Surv.},
  volume    = {57},
  number    = {3},
  pages     = {68:1--68:28},
  year      = {2025},
  month     = mar,
  doi       = {10.1145/3700591}
}

@article{Calvo2025LongEnduranceMRTA,
  author  = {Calvo, {\'A}lvaro and Capit{\'a}n, Jes{\'u}s},
  title   = {Heterogeneous Multirobot Task Allocation for Long-Endurance Missions in Dynamic Scenarios},
  journal = {IEEE Trans. Robot.},
  volume  = {41},
  pages   = {6494--6513},
  year    = {2025},
  doi     = {10.1109/tro.2025.3626651}
}

@article{Zhang2025SemanticSuccessiveRefinement,
  author  = {Zhang, Kexin and Li, Lixin and Lin, Wensheng and Yan, Yuna and Li, Rui and Cheng, Wenchi and Han, Zhu},
  title   = {Semantic Successive Refinement: A Generative {AI}-Aided Semantic Communication Framework},
  journal = {IEEE Trans. Cogn. Commun. Netw.},
  volume  = {11},
  number  = {2},
  pages   = {687--699},
  year    = {2025},
  month   = apr,
  doi     = {10.1109/tccn.2025.3526839}
}

@article{Yan2025AdaptiveSemanticGeneration,
  author  = {Yan, Yuna and Li, Lixin and Zhang, Xin and Lin, Wensheng and Cheng, Wenchi and Han, Zhu},
  title   = {Adaptive Semantic Generation and {NOMA}-Based Interference-Aware Transmission for {6G} Networks},
  journal = {IEEE Trans. Wireless Commun.},
  volume  = {24},
  number  = {3},
  pages   = {2404--2416},
  year    = {2025},
  month   = mar,
  doi     = {10.1109/twc.2024.3520870}
}

@article{Lin2024SemantIC,
  author  = {Lin, Wensheng and Yan, Yuna and Li, Lixin and Han, Zhu and Matsumoto, Tad},
  title   = {{SemantIC}: Semantic Interference Cancellation Toward {6G} Wireless Communications},
  journal = {IEEE Commun. Lett.},
  volume  = {28},
  number  = {8},
  pages   = {1810--1814},
  year    = {2024},
  month   = aug,
  doi     = {10.1109/lcomm.2024.3412973}
}

@article{Lin2024SemanticForwardRelaying,
  author  = {Lin, Wensheng and Yan, Yuna and Li, Lixin and Han, Zhu and Matsumoto, Tad},
  title   = {Semantic-Forward Relaying: A Novel Framework Toward {6G} Cooperative Communications},
  journal = {IEEE Commun. Lett.},
  volume  = {28},
  number  = {3},
  pages   = {518--522},
  year    = {2024},
  month   = mar,
  doi     = {10.1109/lcomm.2024.3352916}
}

@inproceedings{Zhang2026LLSC,
  author    = {Zhang, Kexin and Xu, Dongwei and Lin, Wensheng and Guo, Jinlong and Li, Lixin and Han, Zhu},
  title     = {{LLSC}: End-to-End Image Semantic Communication Framework for Low-Light Scenarios},
  booktitle = {Proc. IEEE Int. Conf. Commun. (ICC)},
  address   = {Glasgow, United Kingdom},
  pages     = {1--6},
  year      = {2026},
  month     = may,
  doi       = {10.1109/icc59461.2026.11588060}
}

@inproceedings{Wang2026FVSF,
  author    = {Wang, Shuaitong and Zhang, Kexin and Qin, Zixuan and Wei, Baoguo and Li, Xu and Lin, Wensheng and Li, Lixin},
  title     = {{FVSF}: Fusion of Multi-Scale Visual and Enhanced Semantic Features for Few-Shot Learning},
  booktitle = {Proc. Int. Conf. Electron., Inf., Commun. (ICEIC)},
  address   = {Macau, China},
  pages     = {1--6},
  year      = {2026},
  month     = jan,
  doi       = {10.1109/iceic69189.2026.11386329}
}

@inproceedings{Huang2025SwinSemantIC,
  author    = {Huang, Yizheng and Lin, Wensheng and Li, Lixin and Han, Zhu},
  title     = {{Swin-SemantIC}: A Transformer-Driven Enhancement Scheme for Semantic Interference Cancellation},
  booktitle = {Proc. IEEE Conf. Cloud Big Data Comput. (CBDCom)},
  address   = {Hakodate, Japan},
  pages     = {69--74},
  year      = {2025},
  month     = oct,
  doi       = {10.1109/cbdcom68404.2025.00017}
}

@article{Wu2025AIEnabledISCCSurvey,
  author  = {Wu, Guofang and He, Yejun and Cao, Xiaowen and Yuen, Chau},
  title   = {{AI}-Enabled Integrated Sensing, Communication, and Computation Survey: Techniques, Status, and Perspectives},
  journal = {IEEE Internet Things J.},
  volume  = {12},
  number  = {16},
  pages   = {32676--32700},
  year    = {2025},
  month   = aug,
  doi     = {10.1109/jiot.2025.3579940}
}

@article{Khoramnejad2025GenerativeAIOptimization,
  author  = {Khoramnejad, Fahime and Hossain, Ekram},
  title   = {Generative {AI} for the Optimization of Next-Generation Wireless Networks: Basics, State-of-the-Art, and Open Challenges},
  journal = {IEEE Commun. Surveys Tuts.},
  volume  = {27},
  number  = {6},
  pages   = {3483--3525},
  year    = {2025},
  month   = dec,
  doi     = {10.1109/comst.2025.3535554}
}

@article{Fu2026GenerativeAITasCom,
  author  = {Fu, Yuzhou and Cheng, Wenchi and Wang, Jingqing and Yin, Liuguo and Zhang, Wei},
  title   = {Generative {AI} Driven Task-Oriented Adaptive Semantic Communications},
  journal = {IEEE Trans. Wireless Commun.},
  volume  = {25},
  pages   = {9078--9093},
  year    = {2026},
  doi     = {10.1109/twc.2025.3644355}
}

@article{Ma2026Condensed360,
  author  = {Ma, Yang and Cheng, Wenchi and Wang, Jingqing and Zhang, Wei and Zhang, Hailin},
  title   = {Condensed Semantic Communication for {360$^{\circ}$} Image Transmission},
  journal = {IEEE Trans. Mobile Comput.},
  volume  = {25},
  number  = {6},
  pages   = {7859--7873},
  year    = {2026},
  month   = jun,
  doi     = {10.1109/tmc.2025.3647481}
}

@article{Fu2024DigitalAnalog,
  author  = {Fu, Yuzhou and Cheng, Wenchi and Zhang, Wei and Wang, Jingqing},
  title   = {Digital-Analog Transmission Framework for Task-Oriented Semantic Communications},
  journal = {IEEE Netw.},
  volume  = {38},
  number  = {6},
  pages   = {81--88},
  year    = {2024},
  month   = nov,
  doi     = {10.1109/mnet.2024.3435848}
}

@article{Fu2024ScalableExtraction,
  author  = {Fu, Yuzhou and Cheng, Wenchi and Zhang, Wei and Wang, Jingqing},
  title   = {Scalable Extraction Based Semantic Communication for {6G} Wireless Networks},
  journal = {IEEE Commun. Mag.},
  volume  = {62},
  number  = {7},
  pages   = {96--102},
  year    = {2024},
  month   = jul,
  doi     = {10.1109/mcom.021.2300269}
}

@article{Wang2026MultiQoSHRLLC,
  author  = {Wang, Jingqing and Cheng, Wenchi and Zhang, Wei},
  title   = {{AI}-Enabled Multi-{QoS} Provisioning for {6G} {HRLLC}},
  journal = {IEEE Commun. Mag.},
  year    = {2026},
  note    = {early access},
  doi     = {10.1109/MCOM.001.2500666}
}

@article{Wang2025AoIResourceAllocation,
  author  = {Wang, Jingqing and Cheng, Wenchi and Zhang, Wei},
  title   = {{AoI}-Aware Resource Allocation for Smart Multi-{QoS} Provisioning},
  journal = {IEEE Syst. J.},
  volume  = {19},
  number  = {1},
  pages   = {305--316},
  year    = {2025},
  month   = mar,
  doi     = {10.1109/jsyst.2024.3519536}
}

@article{Ma2026LargeAIImmersive,
  author  = {Ma, Yang and Cheng, Wenchi and Wang, Jingqing and Zhang, Wei and Zhang, Hailin},
  title   = {Large {AI} Models-Driven Immersive Communication},
  journal = {IEEE Commun. Mag.},
  volume  = {64},
  number  = {5},
  pages   = {132--138},
  year    = {2026},
  month   = may,
  doi     = {10.1109/mcom.001.2500419}
}

@article{Zhang2026MAHEADNet,
  author  = {Zhang, Yixin and Yao, Zhuohui and Cheng, Wenchi and Saad, Walid},
  title   = {{MA-HEAD-Net}: Adaptive Rule-Guided Multi-Agent {DRL} for {AoI} Minimization in {UAV}-Assisted Emergency Networks},
  journal = {IEEE Trans. Cogn. Commun. Netw.},
  volume  = {12},
  pages   = {9962--9978},
  year    = {2026},
  doi     = {10.1109/tccn.2026.3711100}
}

@article{Xiao2026AoGI,
  author  = {Xiao, Yuquan and Du, Qinghe and Cheng, Wenchi and Karagiannidis, George K. and Nallanathan, Arumugam and Guizani, Mohsen and Wang, Jiangzhou},
  title   = {Age of Generative Information ({AoGI}): An Inference-Aware Metric for Freshness Evaluation in Real-Time Computer Vision},
  journal = {IEEE Trans. Mobile Comput.},
  volume  = {25},
  number  = {10},
  pages   = {17165--17178},
  year    = {2026},
  month   = oct,
  doi     = {10.1109/tmc.2026.3689106}
}

@article{Xiao2025StatisticalAoI,
  author  = {Xiao, Yuquan and Du, Qinghe and Karagiannidis, George K.},
  title   = {Statistical Age of Information: A Risk-Aware Metric and Its Applications in Status Updates},
  journal = {IEEE Trans. Wireless Commun.},
  volume  = {24},
  number  = {3},
  pages   = {2325--2340},
  year    = {2025},
  month   = mar,
  doi     = {10.1109/twc.2024.3520324}
}

@article{Xiao2026RFUAVIdentification,
  author  = {Xiao, Yuquan and Du, Qinghe and Zhao, Zixiao and Li, Bin and Tang, Xiao and Zhang, Shijiao and Song, Houbing},
  title   = {{RF}-Based Identification Framework Against Unauthorized {UAV} Networking in Low-Altitude Economy},
  journal = {IEEE Trans. Netw. Sci. Eng.},
  volume  = {13},
  pages   = {6538--6555},
  year    = {2026},
  doi     = {10.1109/tnse.2026.3658563}
}

\end{document}